\documentclass[
preprint2
]{aastex}

\usepackage[fleqn]{amsmath}
\usepackage{graphicx}
\usepackage[FIGTOPCAP]{subfigure}
\usepackage{txfonts}
\usepackage{natbib}
\bibpunct{(}{)}{;}{a}{}{,}
\usepackage[colorlinks=true,citecolor=black]{hyperref}

\usepackage{color}

\newcommand{\vONE}[1]{#1}

\usepackage{stackengine}

\title{
Protostellar Outflows and Radiative Feedback from Massive Stars. \\
II. Feedback, Star Formation Efficiency, and Outflow Broadening
}
\author{Rolf Kuiper}
\affil{
Institute of Astronomy and Astrophysics, University of T\"ubingen, Auf der Morgenstelle 10, D-72076 T\"ubingen, Germany}
\affil{
Max Planck Institute for Astronomy, K\"onigstuhl 17, D-69117 Heidelberg, Germany}
\email{rolf.kuiper@uni-tuebingen.de} 
\and 
\author{Neal J. Turner}
\affil{Jet Propulsion Laboratory, California Institute of Technology, 4800 Oak Grove Drive, Pasadena, CA 91109, USA}
\email{Neal.J.Turner@jpl.nasa.gov}
\and
\author{Harold W.~Yorke}
\affil{Jet Propulsion Laboratory, California Institute of Technology, 4800 Oak Grove Drive, Pasadena, CA 91109, USA}
\email{Harold.W.Yorke@jpl.nasa.gov} 

\begin{abstract}
In this second of a series of radiation-hydrodynamical studies of protostellar outflows and radiative force feedback from massive protostars we perform 2D axially symmetric simulations
to assess the impact of varying
1) when the protostellar outflow starts,
2) the ratio of ejection to accretion rates, and
3) the strength of the wide angle disk wind component.
The star formation efficiency, defined here as the ratio of final stellar mass to initial core mass, is dominantly controlled by radiative forces and the assumed ratio of ejection to accretion rates of the outflow.
Increasing this ratio has three effects: 
First, the protostar grows more slowly and thus has lower luminosity at \vONE{any} given time, lowering the radiative feedback.
Second, the low density bipolar cavity cleared by the outflow is larger, which further diminishes the radiative feedback on disk and core scales.
Third, the higher momentum outflow sweeps up more material from the collapsing envelope, and the potential mass reservoir of the forming star is decreased via entrainment.
The star formation efficiency varies with the ratio of ejection to accretion rates from 50\% in the case of very weak outflows to as low as 20\% for very strong outflows.
At latitudes between the low density bipolar cavity and the high density accretion disk, wide angle disk winds remove some of the gas, which otherwise would be part of the accretion flow onto the disk;
varying the strength of these wide angle disk winds, however, alters the final star formation efficiency by only $\pm6\%$.
For all cases, the opening angle of the bipolar outflow cavity remains below $20\degr$ during early protostellar accretion phases, increasing rapidly up to $65\degr$ at the onset of radiation pressure feedback.
\end{abstract}

\begin{abstract}
We perform two-dimensional axially symmetric radiation-hydrodynamic simulations to assess the impact of outflows and radiative force feedback from massive protostars by varying when the protostellar outflow starts, the ratio of ejection to accretion rates, and the strength of the wide angle disk wind component. 
The star formation efficiency, i.e.~the ratio of final stellar mass to initial core mass, is dominated by radiative forces and the ratio of outflow to accretion rates. 
Increasing this ratio has three effects: 
First, the protostar grows slower with a lower luminosity at any given time, lowering radiative feedback. 
Second, bipolar cavities cleared by the outflow are larger, further diminishing radiative feedback on disk and core scales. 
Third, the higher momentum outflow sweeps up more material from the collapsing envelope, decreasing the protostar's potential mass reservoir via entrainment.
The star formation efficiency varies with the ratio of ejection to accretion rates from 50\% in the case of very weak outflows to as low as 20\% for very strong outflows. 
At latitudes between the low density bipolar cavity and the high density accretion disk, wide angle disk winds remove some of the gas, which otherwise would be part of the accretion flow onto the disk; 
varying the strength of these wide angle disk winds, however, alters the final star formation efficiency by only $\pm 6\%$. 
For all cases, the opening angle of the bipolar outflow cavity remains below $20\degr$ during early protostellar accretion phases, increasing rapidly up to $65\degr$ at the onset of radiation pressure feedback.
\end{abstract}

\keywords{
accretion, accretion disks ---
methods: numerical ---
stars: formation ---
stars: jets ---
stars: massive ---
stars: winds, outflows
\\
\copyright\ 2016. All rights reserved
}

\begin{document}
\maketitle

\section{Introduction}
\label{sect:introduction}
Massive stars impact their natal environment via a variety of feedback effects.
To begin with they inject momentum, mechanical and thermal energy into their surroundings via protostellar jets and outflows and radiation pressure. 
The irradiation and heating also modify the gas' chemical state, an effect not considered further here.
Later, additional mechanical/thermal input comes from 
heating due to photo-ejection of electrons from dust, molecular dissociation and ionization,
stellar winds, and
supernovae.
In this investigation, we address the feedback effects of the protostellar phases only, namely protostellar outflows and radiation pressure.
The impact of protostellar outflows can be subdivided into three components:
First, the redirection of a fraction of the accretion flow into an outflow implies a decrease of the actual stellar accretion rate.
Second, outflows inject outward directed momentum into the infalling gas and hence counteract the stellar gravitational attraction, resulting in a slow-down or even reversal of the infall from the pre-stellar core ({\sl entrainment}).
Third, outflows produce low density bipolar cavities, which in turn alter the effects of radiation pressure during later phases.
In the first article of this series \citep[][hereafter Paper I]{Kuiper:2015p28986}, we labeled these components as
``mass loss feedback'',
``kinematic feedback'', and
``radiative feedback'', respectively.

In Paper I we focussed on the latter of the three protostellar outflow feedback effects.
These previous simulations used a nominal value of the ratio of ejection to accretion rates of only 1\%, minimizing the effects of mass loss feedback and kinematic feedback, but allowing us to investigate how protostellar outflows change the morphology of the stellar environment and how this affects the efficiency of the later radiative feedback phase.
We found that the low density outflow cavities initiate a large scale anisotropy of the thermal radiation field, which 
extends the so-called flashlight effect from the disk out into the core.
The disk's flashlight effect denotes the anisotropy of the thermal radiation field around a forming star due to the high optical depth of its accretion disk \citep{Nakano:1989p497, Yorke:1999p156, Yorke:2002p735, Kuiper:2010p541, Kuiper:2011p21204}.
The core's flashlight effect enables sustained accretion from the core to the disk in an analogous fashion as the disk's flashlight effect enables sustained accretion from the disk to the star.
Further details on the impact of radiation pressure feedback is given in Sect.~\ref{sect:intro_previous}, where we discuss the current investigation in the context of our former simulation studies.

Here, we consider the effects of higher mass loss and vary the ratio of ejection to accretion rates from 1\% up to a maximum value of 50\%.
Additionally, we explore the broad parameter space of outflow configurations by varying its launching time and the strength of a large angle disk wind component.
Such a numerical study necessarily involves considering a broad range of parameter space, see Sect.~\ref{sect:parameterspace} for details.

In general, jets and outflows vary strongly with respect to the ratio of ejection to accretion rates and degree of collimation.
First trends of these variations have been detected, e.g.~collimation seems to weaken with increasing age and stellar mass \citep{Beuther:2005p142}.
The current theoretical and observational understanding of jets and outflows does not allow one to fully trace these parameters back to stellar and/or environmental properties.
Naturally, the launching physics on the smallest scale is the most difficult to obtain from observations, and the launching picture lacks of a detailed theoretical description as well.
A review of observational outflow studies with a focus on the launching physics is given by e.g. \citet{Ray:2007p10756}.
The recent review by \citet{Frank:2014p29566} is structured along spatial scales of jets and outflows and also includes a chapter on the launching scales.
Other observational reviews in this context were presented by \citet{Arce:2007p10514} and \citet{Bally:2008p20721}.
For a description of the theoretical context of jets and outflows from young stars, we refer the reader to the reviews by \citet{Konigl:2000p9442} and \citet{Pudritz:2007p549}.

\subsection{Feedback effects of protostellar outflows}
Outflows are considered an important feedback mechanism which reduces the overall star formation efficiency, especially on spatial scales from pre-stellar cores to accretion disks \citep[see e.g.~review by][and references therein]{Frank:2014p29566}.
\citet{Banerjee:2007p691} and \citet{Hennebelle:2011p5748} followed the collapse of a magnetized pre-stellar core collapse 
to study the self-consistent launching of a bipolar outflow from a forming high mass star, but both studies neglected the radiative feedback of the protostar.
More recent studies \citep{Wang:2010p2487, Cunningham:2011p953, Peters:2014p27736, Federrath:2014p28884, Kuiper:2015p28986} make use of subgrid modules for protostellar outflow feedback.
\citet{Wang:2010p2487} studied the interaction of outflows and the ambient large scale magnetic field in a cluster-formation simulation, which leads to a kind of self-regulation of the overall collapse in the sense that higher accretion rates are coupled to higher feedback efficiencies.
In the context of cluster formation, \citet{Peters:2014p27736} presented a study of the combined feedback of multiple low mass outflows as an alternative explanation for high mass loss rates and momentum feedback, if the associated protostars are forming with nearly the same direction of their angular momentum vectors, e.g.~as a consequence of a global common rotation of the cluster-forming gas.
The overall reduction of the star formation efficiency due to the outflows' kinematic feedback during cluster formation was presented in \citet{Federrath:2014p28884}, focussing on low mass stars as well.

Simulations of higher mass protostars, including protostellar outflows and radiative feedback were presented in \citet{Cunningham:2011p953}.
They report a reduction of radiative flux in the equatorial plane perpendicular to the outflow due to the low density outflow cavities, a finding, which we extended in Paper I to the super-Eddington regime, including strong radiative forces by massive protostars.
Studies of the combined effects of protostellar outflows and radiative forces remain limited.
Here, we expand on our earlier work by covering the parameter space of outflow injection, including different strengths of the total outflow and of a disk wind.

One of the key aspects of our numerical studies is the long evolutionary timescales covered by the simulations. 
Our simulations are stopped after the total depletion of the mass reservoir due to feedback; this process takes up to 10 free-fall times of the original pre-stellar core.
The modeling of the full stellar accretion and feedback phase allows a quantitative determination of the overall feedback efficiencies in terms of total mass loss.
By contrast, the numerical simulations cited above cover timescales of about 
one free-fall time or less \citep{Banerjee:2007p691, Cunningham:2011p953, Peters:2014p27736},
up to 2 free-fall times only in simulations without radiation transport \citep{Hennebelle:2011p5748, Federrath:2014p28884},
and up to a maximum of 5 free-fall times only in the simulation by \citet{Wang:2010p2487}.

From an observational perspective of large cluster-scales, 
magnetic-field-regulated accretion \citep[e.g.][]{Vlemmings:2010p10559} 
and outflow regulated feedback \citep[e.g.][]{Nakamura:2014p28897}
seem to be likely.
Molecular outflows from protostars may even contribute to the final mass depletion of the stellar surroundings \citep[see e.g.][]{Shepherd:2004p31878}.
On these scales, outflows may also contribute to the replenishment of turbulence within the star-forming region \citep[see review by][and references therein]{MacLow:2004p350}.
\vONE{
However the importance of this contribution is being questioned and observations seem to produce contradictory results \citep{Arce:2011p32946, DrabekMaunder:2016p32992}.
}

\subsection{Outline of our preceding studies}
\label{sect:intro_previous}

In this section, we embed the simulation series into (our) preceding simulation studies regarding massive stars in the super-Eddington regime.
Massive stars can become so luminous that their stellar radiative force on the directly illuminated surrounding gas and dust exceeds their gravitational attraction.
For spherically symmetric accretion flows, this feedback causes the so-called radiation pressure problem in the formation of massive stars \citep{Kahn:1974p799, Yorke:1977p376}, which sets a stellar upper mass limit of about $40 \mbox{ M}_\odot$.
Based on the studies by \citet{Nakano:1989p497} and \citet{Yorke:1999p156}, it is expected that this radiation pressure problem is reduced in the presence of anisotropic thermal radiation fields, which are naturally generated by a massive accretion disk forming around the protostar, also known as the ``disk's flashlight effect''.
In subsequent numerical studies of the anisotropic radiation pressure in the formation of massive stars \citep{Yorke:2002p735, Krumholz:2009p687} the highest mass of the forming protostars was still limited to $M_* \lesssim 43 \mbox{ M}_\odot$, only marginally above the 1D limit.
The formation of massive stars up to $M_* \lesssim 140 \mbox{ M}_\odot$ was demonstrated in \citet{Kuiper:2010p541}, utilizing a more accurate treatment of the radiation transport and higher resolution of the forming accretion disk.
The force analysis of the simulation data shows that the radiation pressure problem is indeed circumvented by the flashlight effect.

Later on, these numerical models were enhanced from axially symmetric configurations to three-dimen\-sional simulations, investigating the non-axially symmetric morphology of the accretion disk and the radiation-pressure-dominated outflow region \citep{Kuiper:2011p21204}.
Self-gravity in the massive accretion disk generates spiral arms, which transport angular momentum outward, while gas is transported inward.
A comparison of the resulting accretion with the results of axially symmetric simulations with different strengths of the so-called $\alpha$-shear viscosity, reveals the self-gravity as a very efficient driver of the disk's accretion towards the protostar.
The disk accretion rates driven by the self-gravity were found to be high enough to overcome the diminished radiation pressure in the disk.

Although direct observations of massive accretion disks at these early stages are extremely difficult, very recent ALMA observations were able to reveal Keplerian-like rotation profiles around a high mass O7 star \citep{Johnston:2015p32366}.

The inner gaseous disk surrounding a massive protostar inside the dust sublimation zone further contributes to the disk's flashlight effect and at the same time shields the large scale collapsing environment from the direct stellar irradiation, allowing for sustained envelope-to-disk accretion \citep{Kuiper:2013p17358}.
These results were obtained by including the gas opacities from \citet{Helling:2000p1117}, also used in the protoplanetary disk models by \citet{Semenov:2003p79}.

We studied the impact that stellar evolution has on the environment and vice versa in \citet{Kuiper:2013p19987}.
Whereas on long timescales, this interplay seems to have only limited consequence on the final mass, on timescales $< 5$~kyr, the interplay of stellar evolution and accretion results in variable accretion and accretion bursts.
More interestingly, different initial conditions of the cloud collapse yield a broad range of the protostellar bloating epoch.
In the different runs, the protostar reaches the zero-age main-sequence between $20 \mbox{ M}_\odot$ and $40 \mbox{ M}_\odot$.
This point in evolution is of special interest, because it denotes the beginning of strong ionization feedback of the forming massive star, creating an expanding HII region around it.

The reduction of the radiative impact on the disk accretion flow due to low density outflow cavities has been studied in both purely static \citep{Krumholz:2005p406} and hydrodynamic models \citep[][Paper I]{Cunningham:2011p953
}.
In Paper I we demonstrate that protostellar outflows can enlarge the anisotropy out to core scales.
This is the basis for what we call the ``core's flashlight'' effect, i.e.~the core-scale anisotropy, which shields
large portions of the core from intense thermal radiation,
allowing mass to accrete from the collapsing envelope onto the accretion disk in an analogous manner to the disk's flashlight effect, which allows mass to accrete from the disk onto the (proto)star.
With the present investigation we further elaborate on these previous studies.

Methodologically, the outcome of radiation hydro\-dynamics simulations in the super-Eddington regime depends crucially on the accuracy of the method used to compute the stellar radiative feedback \citep{Kuiper:2012p1151}.
In the past, most large scale star formation simulations were limited to the gray flux-limited diffusion approximation, which generally provides an accurate estimate of the dust temperature but which yields a very poor estimate of the radiative force.
Recently,
\citet{Klassen:2014p27924} introduced an implementation of the hybrid radiation transport scheme of \citet{Kuiper:2010p586} into the adaptive grid code FLASH \citep{Fryxell:2000p768, Dubey:2009p32315}, 
\citet{Harries:2015p32122} added a hydrodynamics module for the Monte Carlo radiative transfer code TORUS, and
\citet{Buntemeyer:2016p32167} independently implemented a radiation transport module for FLASH using characteristics on the adaptive grid.
First simulations carried out with these tools \citep{Harries:2014p28908, Klassen:2016p32943} address the accuracy of radiative forces in the formation of a massive protostar as discussed in \citet{Kuiper:2012p1151} and confirm these earlier results.
Furthermore, the simulations by \citet{Klassen:2016p32943} and \citet{Kuiper:2011p21204} agree in identifying the self-gravity of the massive accretion disks as a sufficiently strong driver of angular momentum transport, and hence mass accretion against the diminished radiative force.

\section{Parameter space}
\label{sect:parameterspace}
In this section, we summarize how we have constrained the parameter space and reasons for our choice of parameters.
These constraints are, whenever possible, based on observations of jets and outflows from massive protostars as well as numerical studies with respect to their launching and early evolution on smaller scales.
Although we are studying the impact of jets and outflows from massive stars, knowledge of the outflow properties from low mass stars might help constrain the parameterization of massive star outflows, if we accept the idea that the outflow physics are fundamentally the same (which is not necessarily the case).
Hence, if parameter constraints are only available for lower mass protostars, we try to extrapolate the massive star's parameter constraints from them.

\subsection{Ratio of ejection to accretion rates}
Although the details are under debate, observational and theoretical studies support the basic idea that the ejection rate of protostellar outflows is (strongly) correlated to the stellar accretion rate.
\citet{Benisty:2010p31762}, for example, presented an observed correlation between accretion outburst and mass loss for an embedded Herbig Be star.
Further examples of observations of disk-jet or disk-outflow systems around intermediate and high mass protostars are given in \citet{Ellerbroek:2011p5471} and \citet{Johnston:2015p32366}.
A rough sketch of the magnetic field and velocity structure of the accretion flow and outflow around a forming high mass star was recently presented in \citet{Sanna:2015p32584}.

High mass protostars are usually characterized by high accretion rates and show high mass loss rates of up to several $10^{-3} \mbox{ M}_\odot \mbox{ yr}^{-1}$ as well \citep{Beuther:2002p3046, Beuther:2002p3040, Wu:2004p20604, Wu:2005p772, Bally:2008p20721}.
These massive protostars are born in very dense environments, and their environment is characterized by complex morphologies such as multiple outflows, possibly as a result of the high degree of multiplicity among massive stars \citep[e.g.][]{Zinnecker:2007p363}.
At the high mass stellar spectrum, ionization and radiative forces also play a crucial role for the dynamics of their environment.
Disentangling these effects from the MHD-driven outflow components is difficult, hence, an observational determination of mass loss rates due to outflows alone, i.e. without contributions from radiative and mechanical feedback, remains uncertain.
Hence, whereas accretion and intrinsic outflow rates can be determined for lower mass protostars, they are currently inaccessible for high mass stars, i.e.\ the observed ratios of the ejection to accretion rates are based on very indirect measurements and are quite uncertain for higher mass stars \citep[see e.g.][]{Beuther:2013p22520}. 

In the following, we estimate the physically interesting range of the ratio of ejection to accretion rates based on observations from low  and intermediate mass stars.
\citet{Hartigan:1995p31684} give a mean ratio of ejection to accretion rates of roughly 1\% for low mass classical T Tauri stars (but the accretion and ejection rates vary by about two orders of magnitude and their ratio can be significantly higher than 1\%).
More recently, a typical value quoted for the ejection to accretion ratio of low mass classical T Tauri stars is about 10\% \citep[e.g.][and references therein]{Cabrit:2007p31638, Ray:2007p10756, Frank:2014p29566}.
For intermediate mass young stars, observed accretion rates \citep{Calvet:2004p31661} and mass loss rates \citep{AgraAmboage:2009p31644} yield ratios of ejection to accretion rates in the wide range from 2\% to 40\%.
Another compilation of observed ratios of ejection to accretion rates (using a variety of different techniques) from low to intermediate mass stars is shown in \citet{Ellerbroek:2013p31779}, Fig.~14.

Numerical models of the MHD physics including the disk and jet system were presented by e.g.\ \citet{Tomisaka:2001p28345} and \citet{Sheikhnezami:2012p21519} and their calculated ratios of ejection to accretion rates vary from 10\% to even 70\%.
Furthermore, the ratio of ejection to accretion rates might also influence other jet and outflow parameters such as the collimation \citep[see][Sect.~4.4.2]{Fendt:2009p17761}, i.e.~the parameters of the subgrid module are not fully independently of each other.

Observations also suggest a time-dependent accretion and/or ejection variability, see e.g.~the early textbook example of a pulsed jet by \citet{Zinnecker:1998p32379}. This may be the result of a varying accretion rate or a variation of the ratio of outflow to accretion rates.
Here, for simplicity and in order to reduce the number of free parameters, we assume constant fixed rates of ejection to accretion rates in our simulations with values ranging from 1\% up to 50\%.

\subsection{Collimation and large angle disk winds}
In general, observed jets have wide opening angles near to their sources, which are collimated on larger scales typically of the size of their corresponding accretion disks.
The degree of collimation for non-embedded classical T Tauri star outflows lies in the range of 10$^\circ$ to 30$^\circ$ on small scales close to the launching region and collimate afterwards up to the 100 AU scale \citep[see e.g.\ the review by][in particular Fig.~2]{Ray:2007p10756}.
Numerical simulations, which focus on the small scales of the detailed accretion launching interaction at the magnetosphere, e.g.~by \citet{Fendt:2009p17761}, support this self-collimating effect of magneto-centrifugally launched jets.
Comparing their synthetic maps with observations of jets from young stellar objects,
\citet{Stute:2010p31695} conclude that models with launching radii between 0.1 to 1~AU match the observed objects.
Because the numerical size of the sink cell radius of 10~AU used in our simulations is larger than the typical sizes of jet-launching regions, we cannot resolve this initial collimation. 
Nevertheless, the interaction of the outflow with the infalling envelope outside of 10~AU quickly results in a collimated morphology, as presented in Fig.~\ref{fig:snapshots}, panel c.

\citet{Beuther:2005p142} find that the opening angle of outflows generally increase with time.
This could be due to intrinsic processes, e.g.~changes to the MHD launching, or a result of line-driven winds \citep{Vaidya:2011p14600}. 
Rather than trying to model these poorly understood intrinsic processes, we have chosen to keep the angular dependence of the injected material constant during the simulation, while its absolute value varies linearly with the accretion rate. 
We nonetheless find that the outflow's opening angle increases with time in our simulations, due to the weakening of the interaction of infalling envelope with the outflow as the envelope becomes depleted (see
the time series snapshots in Fig.~\ref{fig:snapshots} discussed in Sect.~\ref{sect:broadening}).

Observed opening angles of individual low mass young stellar objects span a broad range from 20$^\circ$ at early phases up to 160$^\circ$ during late phases \citep[see][and references therein]{Frank:2014p29566}.
Moreover, jet and outflow observations typically show an angular dependence of the velocity field consisting of a high velocity collimated jet with decreasing velocity to the wings, often referred to as a ``wide angle (disk) wind''.
To enable studies of the effect of distortion of the large scale envelope by the jet as well as the impact of a large angle disk wind or molecular outflow, we vary the strength of the disk wind component separately from the jet-like component described above.
This variation affects the impact of the outflow on the disk-feeding envelope, such as envelope to disk accretion rates and time evolution of the opening angle of the cleared polar cavity.

\section{The numerical model}
\label{sect:methods}
The parameter space described above extends our initial treatment in Paper I of the dynamical effect of protostellar outflows on the radiative feedback phase of super-Eddington massive stars.
Other numerical parameters and the code itself were not modified.
For a detailed description of the sub-grid module for protostellar outflow feedback, we refer to Paper I.
The overall numerical framework of the star formation code, including e.g.~the self-gravity solver and the disk's shear viscosity prescription, is presented in \citet{Kuiper:2010p541, Kuiper:2011p21204}.

The general numerical algorithm of the radiation-hydro\-dyna\-mics solver is presented in \citet{Kuiper:2010p586}.
This hybrid radiation transport scheme is much more accurate than gray flux-limited diffusion;
its improved accuracy was shown for the case of stellar radiative forces in bipolar cavities \citep{Kuiper:2012p1151, Owen:2014p32582} as well as for the temperature structure of irradiated accretion disks \citep{Kuiper:2013p19458}.

Stellar evolution is treated via interpolating the stellar evolutionary tracks for accreting high mass stars by \citet{Hosokawa:2009p23} in terms of their dependency on stellar mass and accretion rate.
Simultaneous evolution of the (proto)star within its collapsing environment yields variability of the accretion rates on timescales $< 5$~kyr but results in a very similar long-term evolution \citep{Kuiper:2013p19987}.

Dust opacities from \citet{Laor:1993p358} are used; the
gas opacity is set to a constant value of $\kappa_\mathrm{gas} = 0.01 \mbox{ cm}^2 \mbox{ g}^{-1}$.
Higher values of the mean gas opacity, as suggested in \citet{Malygin:2014p27633}, will likely yield additional shielding of stellar radiative feedback \citep{Kuiper:2013p17358}.

\section{Initial Conditions}
\label{sect:setup}
The initial conditions consist of a spherically symmetric, density-peaked ($\rho \sim r^{-3/2}$)
dusty molecular core ($M_\mathrm{core} = 100 \mbox{ M}_\odot$)
at uniform temperature ($T_\mathrm{core} = 20$~K) in axially symmetric solid body rotation ($\Omega = 2 \times 10^{-13}$~s$^{-1}$).
The radiation-hydrodynamics is solved assuming axial and midplane symmetry.
We conduct three series of simulations to investigate the impact of 
the time the protostellar outflow begins,
its ratio of ejection to accretion rates, and
the strength of the large angle disk wind component.

We first perform a ``fiducial run'' with the following parameters:
The outflow is started when the star attains the mass $M_\mathrm{launch} = 8 \mbox{ M}_\odot$;
the ratio of ejection to accretion rates is held constant at $f_\mathrm{ejec-acc} = 20\%$; and
a flattening parameter value of 
$\theta_0 = 0.01$ 
is chosen to parameterize the disk wind component.
The flattening parameter specifies the angular distribution $f(\theta)$ of the injected outflow following \citet{Matzner:1999p13489}:
\begin{equation}
\label{eq:angularweighting}
  f(\theta) = \left\{\ln \left(2/\theta_0 \right) \left[ \sin^2
  (\theta) + \theta_0^2 \right] \right\}^{-1}.
\end{equation}

This default configuration is varied in ten additional simulations.
The stellar mass at which outflow starts is varied from $M_\mathrm{launch}= 2 \mbox{ M}_\odot$ to $16 \mbox{ M}_\odot$.
The ratio of ejection to accretion rates is varied from $f_\mathrm{ejec-acc} = 1\%$ to 50\%.
The parameter $f_\mathrm{ejec-acc}$ is defined so that the outflow mass loss rate is 
$f_\mathrm{ejec-acc} \times \dot{M}_\mathrm{acc, d}$
and the rate of mass growth of the star is
$(1-f_\mathrm{ejec-acc}) \times \dot{M}_\mathrm{acc, d}$, 
where $\dot{M}_\mathrm{acc, d}$ is the calculated mass accreted by the central sink.
The flattening parameter, which modifies the disk-wind component at large angles, is varied from $\theta_0 = 1/300$ to $1/30$.

An overview of these run-time parameters is given in Table~\ref{tab:run-table}.

\begin{table}[htbp]
\begin{center}
\begin{tabular}{l | c c c}
Run label & 
$M_\mathrm{launch}~[\mbox{M}_\odot]$ & 
$f_\mathrm{ejec-acc}$ & 
$\theta_0$
\\
\hline
L-2		  	& 2.0 & 0.2	& $1/100$	\\
L-4		  	& 4.0 & 0.2	& $1/100$	\\
{\bf L-8}		& 8.0 & 0.2	& $1/100$	\\
L-16		  	& 16.0 & 0.2	& $1/100$	\\
\hline
A-0.01 		& 8.0 & 0.01	& $1/100$	\\
A-0.1	  	& 8.0 & 0.1	& $1/100$	\\
{\bf A-0.2}	  	& 8.0 & 0.2	& $1/100$	\\
A-0.3	  	& 8.0 & 0.3	& $1/100$	\\
A-0.4	  	& 8.0 & 0.4	& $1/100$	\\
A-0.5	  	& 8.0 & 0.5	& $1/100$	\\
\hline
D-30		  	& 8.0 & 0.2	& $1/30 $	\\
{\bf D-100}  	& 8.0 & 0.2	& $1/100$	\\
D-300	  	& 8.0 & 0.2	& $1/300$	
\end{tabular}
\end{center}
\caption{ 
Overview of parameters used in the three simulation series. 
The first column specifies the run label.
The three columns to the right refer to
the stellar mass $M_\mathrm{launch}$ at which the protostellar outflow is first launched, 
the ratio of ejection to accretion rates $f_\mathrm{ejec-acc}$, and
the strength of the disk-wind component of the outflow in terms of the flattening parameter $\theta_0$.
Note that the labels ``L-8'', ``A-0.2'', and ``D-100'' denote an identical configuration, meaning a singular run, as they refer to our fiducial  simulation (bold text).
}
\label{tab:run-table}
\end{table}

\section{Results}
\label{sect:results}
Many of the basic features common to all the simulations are shown in Fig.~\ref{fig:snapshots} depicting the fiducial run.
Starting from a gravity-dominated molecular core in slow solid-body rotation, global collapse begins with largely radial infall (Fig.~\ref{fig:snapshot10}), leading to the formation of a protostar in the central sink cell.
A largely centrifugally supported circumstellar accretion disk soon forms (Fig.~\ref{fig:snapshot25}), which launches a bipolar outflow -- injected via a subgrid module.
Here, within the framework of assumed axial symmetry, accretion flow through the disk is driven by a sub-grid viscosity module, designed to mimic the accretion naturally caused by spiral arms within a self-gravitating circumstellar disk \citep[c.f.][]{Kuiper:2011p21204}.
The low density cavity above and below the disk -- created by the protostellar outflow -- enhances the anisotropy of optical depth, and, hence, diminishes the radiative feedback on gas at lower latitudes via the so-called core's flashlight effect, 
a feature discussed in detail in Paper I, where simulation models with and without protostellar outflow feedback are compared.

\begin{figure*}[p]
\vspace{-30mm}
\begin{center}

\setcounter{figure}{1}
\hspace{5mm}
\footnotesize
\stackunder[5pt]{
Gas temperature (K)
}{
\includegraphics[width=0.4\textwidth]{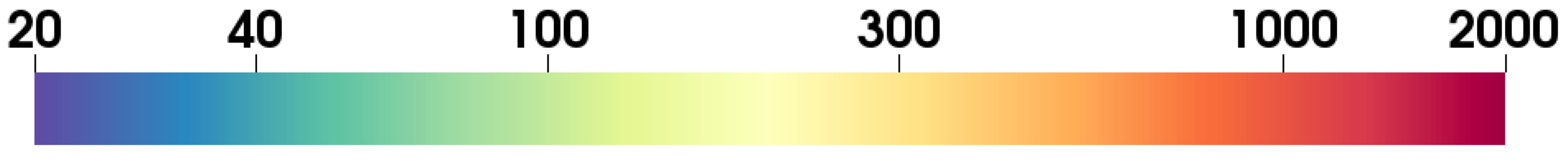}
}
\hspace{7mm}
\stackunder[5pt]{
Gas mass density (g cm$^{-3}$)
}{
\includegraphics[width=0.4\textwidth]{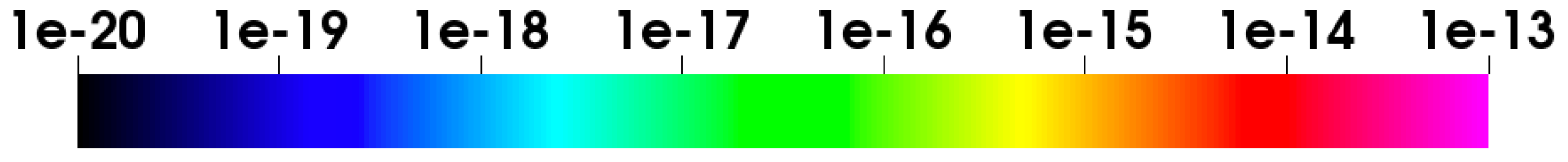}
}
\vspace{5mm}

\subfigure[10 kyr ($M_* = 3.3 \mbox{ M}_\odot$)]{
\label{fig:snapshot10}
\includegraphics[width=0.4\textwidth]{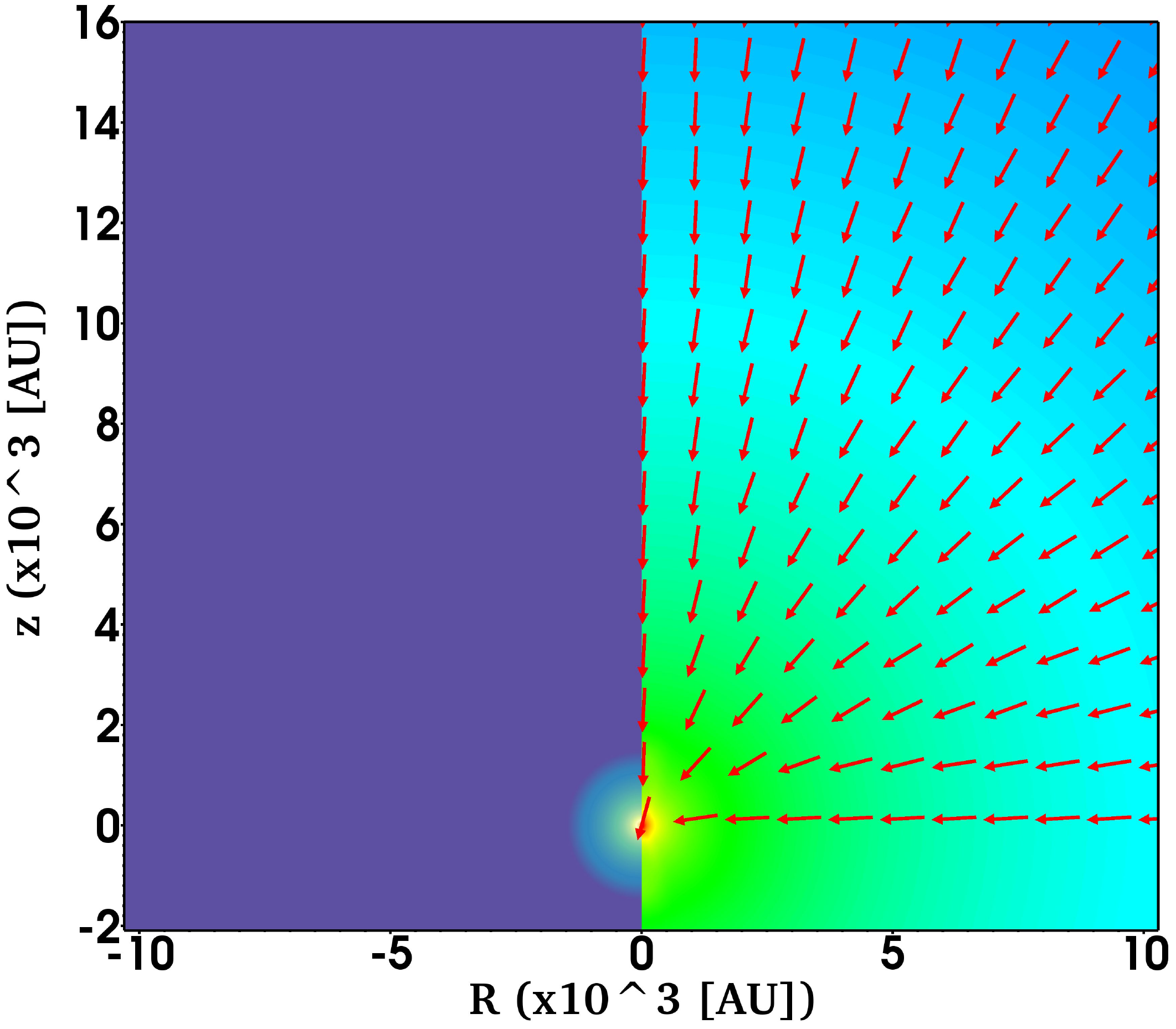}
}
\hspace{5mm}
\subfigure[25 kyr ($M_* = 14.9 \mbox{ M}_\odot$)]{
\label{fig:snapshot25}
\includegraphics[width=0.4\textwidth]{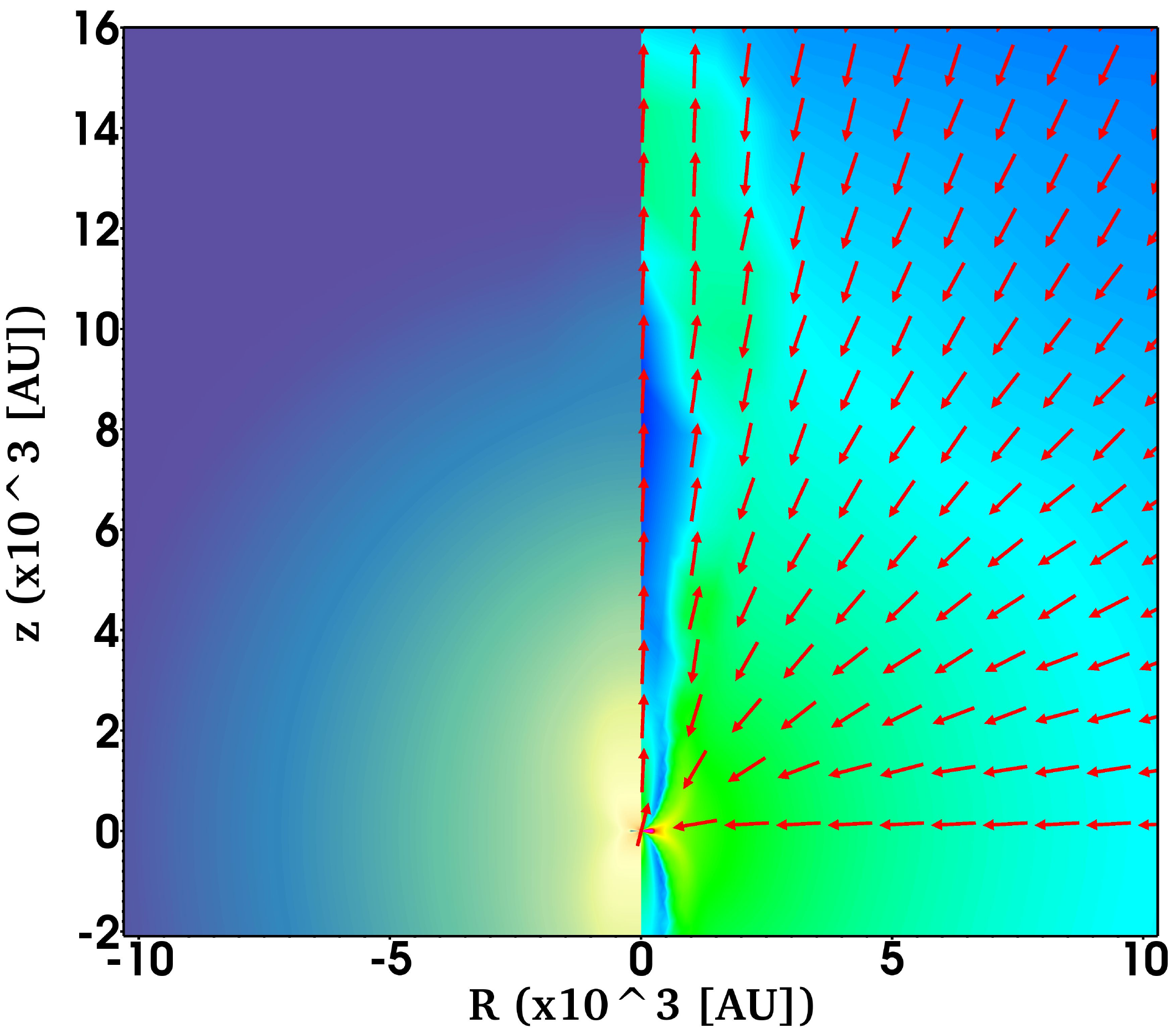}
}
\vspace{5mm}

\subfigure[35 kyr ($M_* = 18.9 \mbox{ M}_\odot$)]{
\label{fig:snapshot35}
\includegraphics[width=0.4\textwidth]{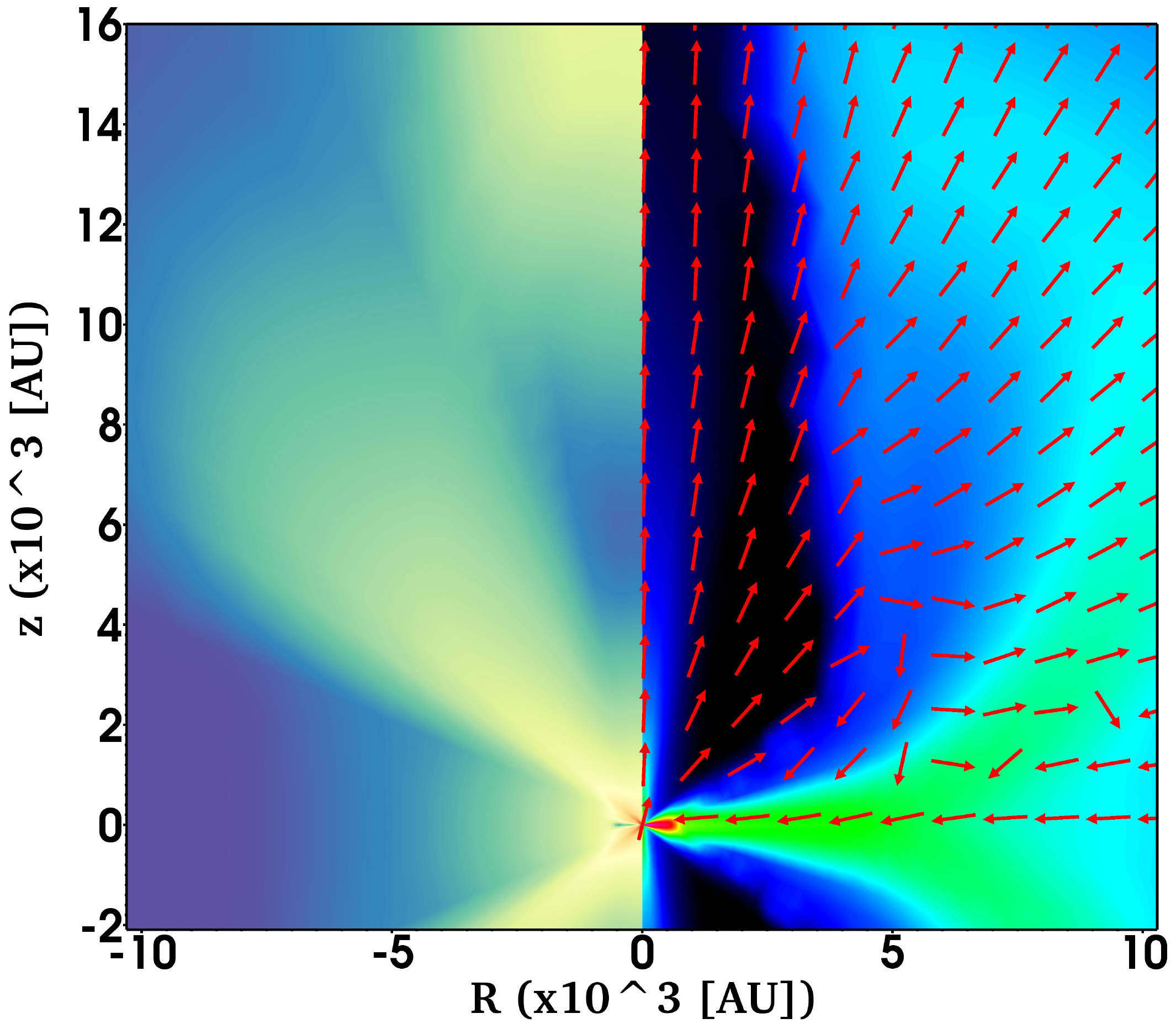}
}
\hspace{5mm}
\subfigure[50 kyr ($M_* = 21.6 \mbox{ M}_\odot$)]{
\label{fig:snapshot50}
\includegraphics[width=0.4\textwidth]{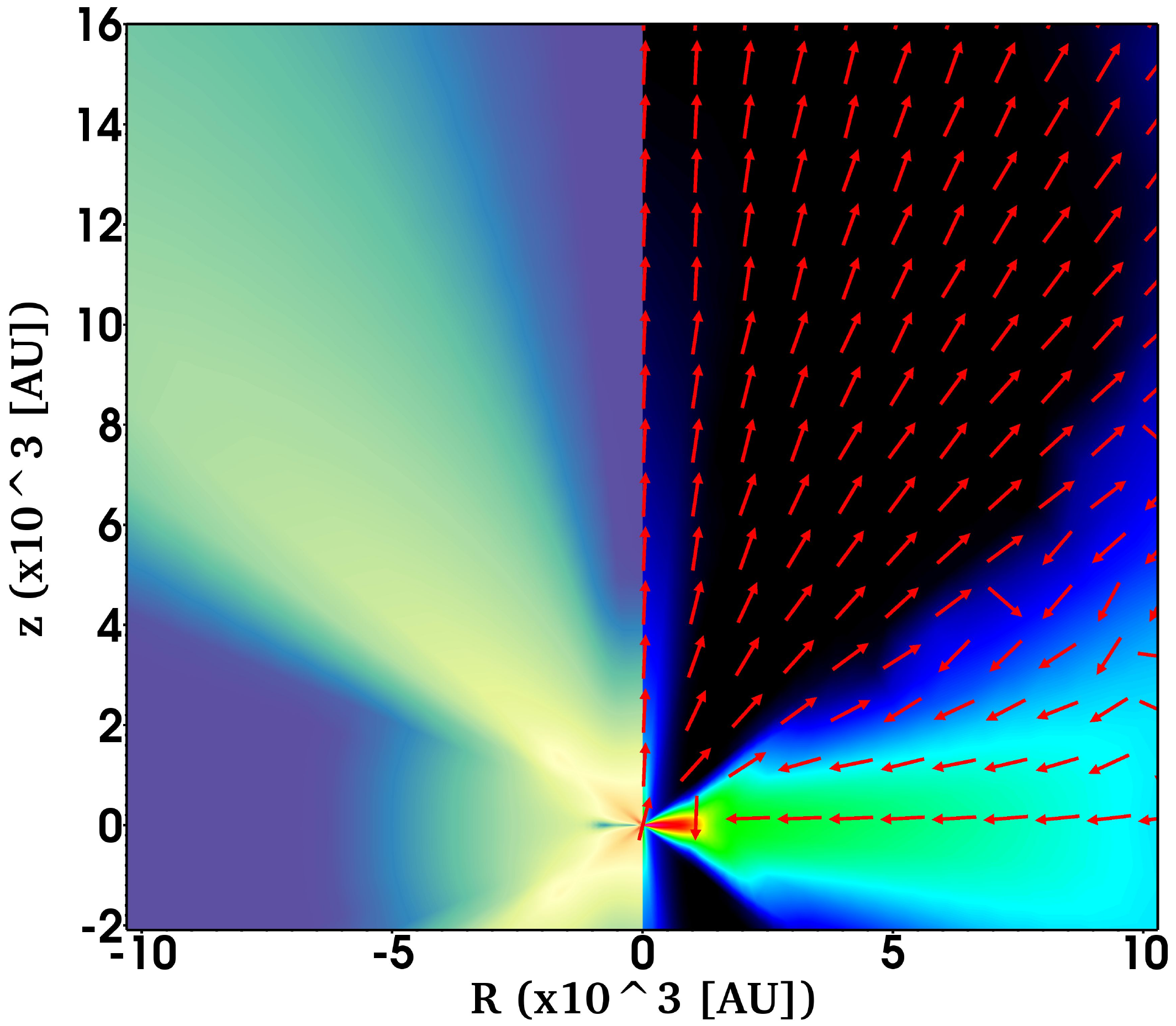}
}
\vspace{5mm}

\subfigure[100 kyr ($M_* = 26.4 \mbox{ M}_\odot$)]{
\label{fig:snapshot100}
\includegraphics[width=0.4\textwidth]{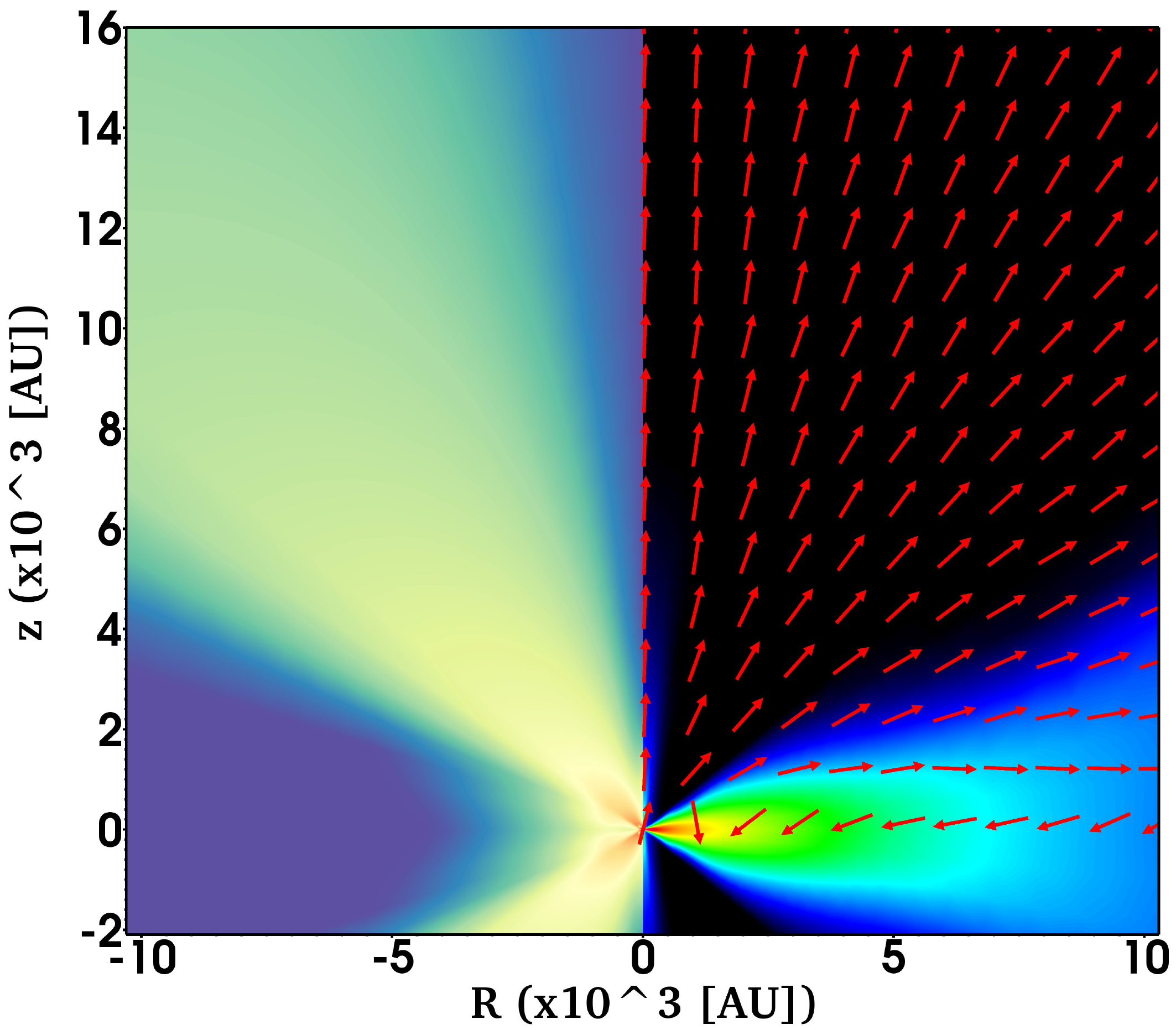}
}
\hspace{5mm}
\subfigure[400 kyr ($M_* = 31.0 \mbox{ M}_\odot$)]{
\label{fig:snapshot400}
\includegraphics[width=0.4\textwidth]{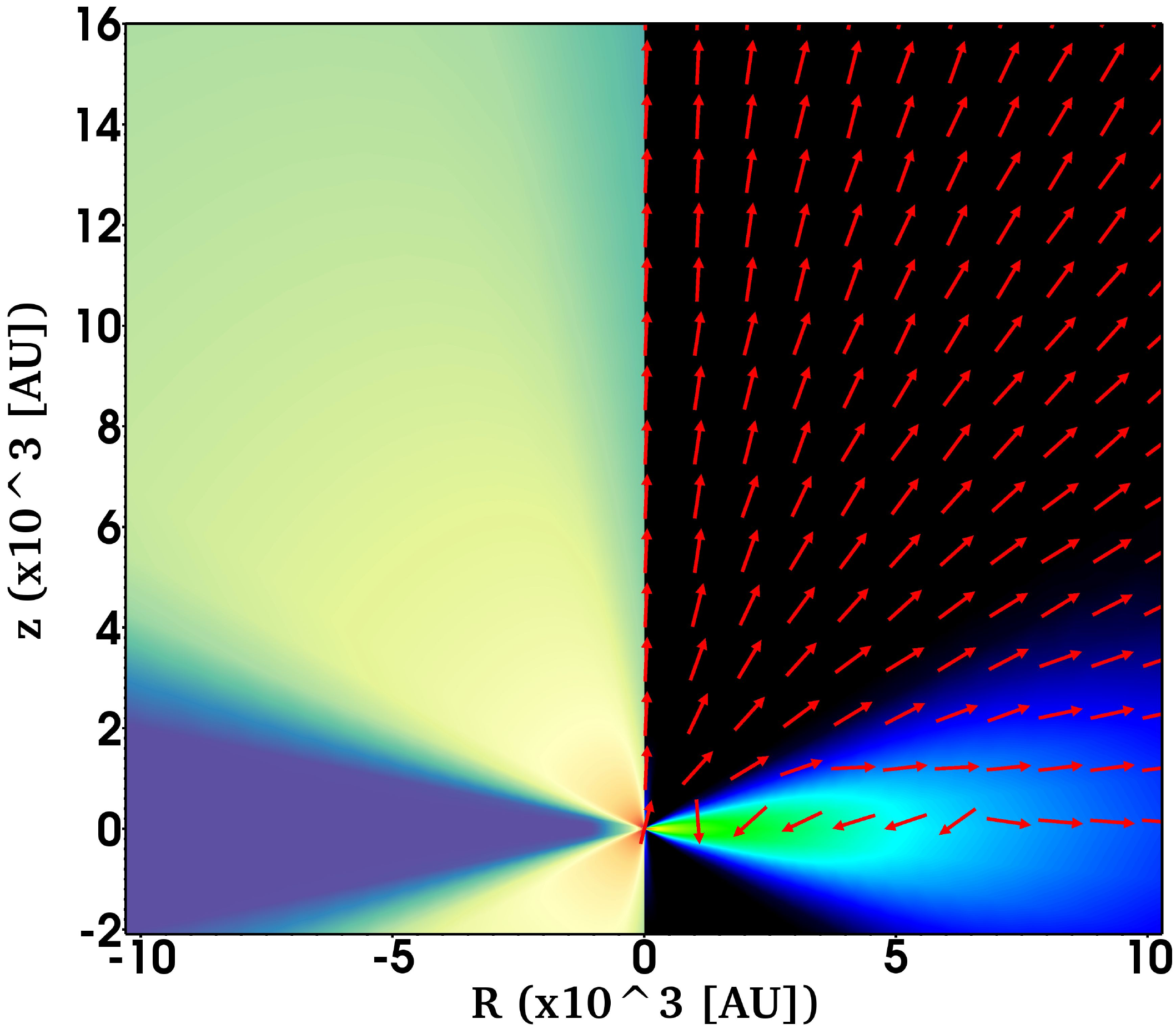}
}

\setcounter{figure}{0}

\caption{
Density (right) and temperature (left) distributions in the central portion of the computational zone 
are shown for different evolutionary times of the fiducial run.
Red arrows on the right show the direction of gas velocity.
}
\label{fig:snapshots}

\end{center}
\end{figure*}

When the stellar radiative force is comparable to the force of gravity ($M_* \gtrapprox 20 \mbox{ M}_\odot$), the low density cavity begins to widen (Fig.~\ref{fig:snapshot35}).
Fig.~\ref{fig:snapshot35} also reveals several warm regions (yellow color on the left side of the image), where the outflow collides with ambient gas.
Although the infall is reversed over ever larger fractions of the volume around the symmetry axis (Fig.~\ref{fig:snapshot50}),
the disk still accretes gas at high rates from latitudes up to $\approx 40\degr$ above the midplane.
The red-colored region in Fig.~\ref{fig:snapshot50} depicts a centrifugally supported circumstellar disk with slow, subsonic inflow;
the flattened green structure is both rotating and in infall.
These rotating, infalling flattened structures represent a transition region between the gravitationally dominated collapsing envelope and the centrifugally supported circumstellar disk.
Following \citet{Beltran:2005p264,Beltran:2011p386}, who observe similar structures, we refer to this region as the rotating torus.

At later times, feeding of the accretion disk from large scales occurs only through the shadowed region behind the disk (Fig.~\ref{fig:snapshot100}).
Fig.~\ref{fig:snapshot400} shows the end of the envelope-to-disk accretion phase.
Accretion from core to disk is lower than loss from the disk into the star and outflow, so the disk's mass declines.
As a result, the disk becomes less opaque and its flashlight effect weakens so that the now more nearly isotropic radiation force halts stellar accretion and removes the remnant gas mass from the stellar vicinity.
For a more detailed investigation of the disk's flashlight effect, which allows a forming massive star to circumvent the so-called radiation pressure barrier, we refer the interested reader to \citet{Kuiper:2010p541}.
We stop the simulations when almost no gas remains in the computational domain; it has either been accreted onto the central star or blown away by the outflow and radiative forces.
These simulations cover the full accretion and feedback phases of the forming massive star over timescales of $\gtrapprox 500$~kyr, corresponding to 10 free-fall times of the parent core.

In the following, we first address each of the simulation series performed -- corresponding to the variation of a single parameter of the underlying outflow properties -- in independent sections.
We then discuss the broadening of the outflow cavities in time for all simulations performed in a separate section.

\subsection{Launching the Outflow  -- $M_{\rm launch}$ Dependence}
Protostellar outflows are assumed to be launched once the magneto-centrifugal acceleration overcomes the gravitational attraction by the protostar.
Both forces -- magnetic and centrifugal -- will strongly increase with the formation of an accretion disk.
It is still a matter of debate, even in the field of low mass star formation, whether
a) the wind-up of the large scale magnetic field after the formation of a differentially rotating accretion disk {\em results} in the launching of a protostellar outflow, or
b) the magnetic fields suppress the early formation of circumstellar disks by efficiently transporting angular momentum to larger radii, see \citet{Frank:2014p29566} and references therein.

\begin{figure}[htbp]
\begin{center}
\includegraphics[width=0.49\textwidth]{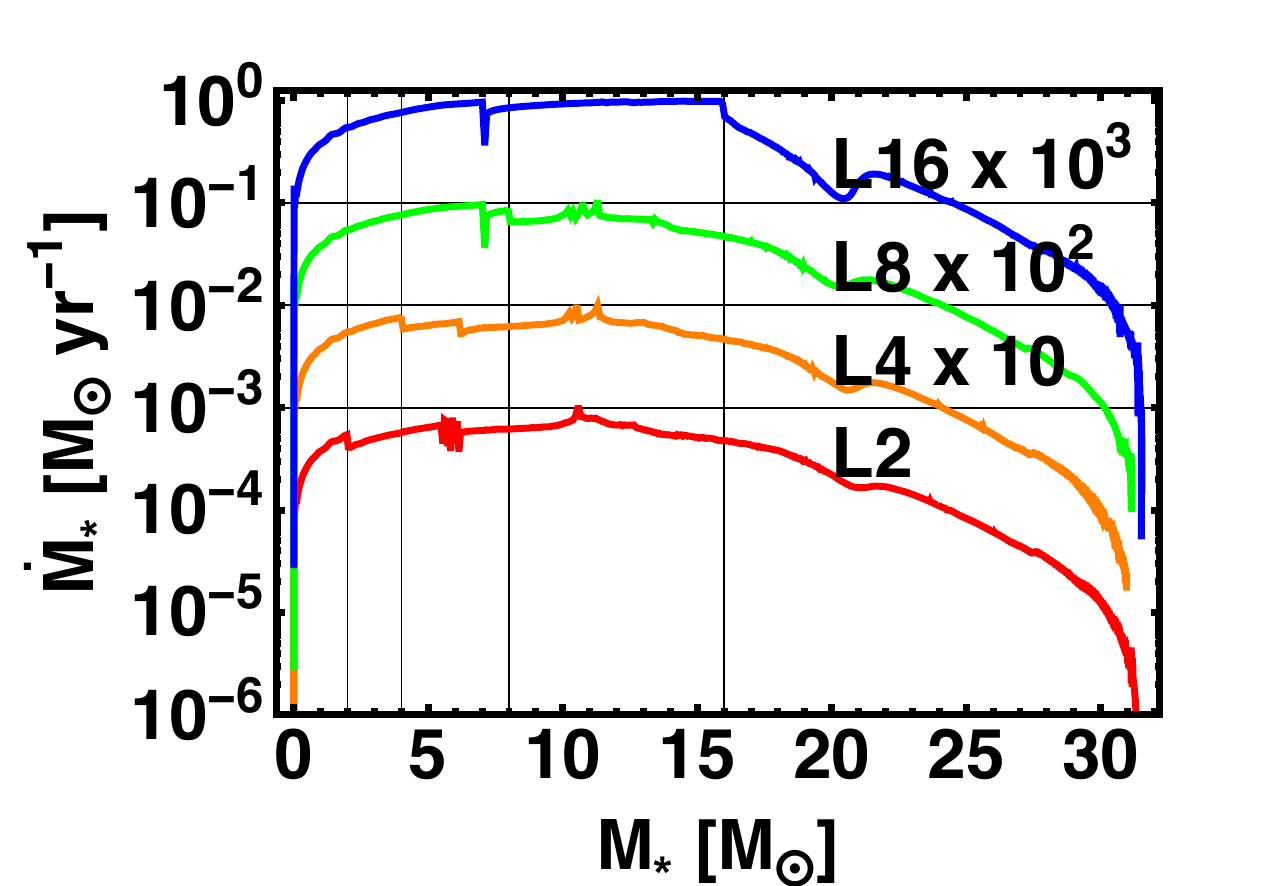}
\end{center}
\caption{
Accretion rate as a function of stellar mass for four different values of the stellar mass $M_\mathrm{launch}$ at which the outflow is launched.
The red, orange, green, and blue curves correspond to $M_\mathrm{launch} = 2, 4, 8, \mbox{ and } 16 \mbox{ M}_\odot$, i.e. cases L2, L4, L8, and L16, respectively.
\vONE{
Because of the fact that these lines would otherwise simply overlap each other, we shifted the resulting lines by one order of magnitude in accretion rate with respect to each other (as labeled).
}
Vertical lines show the different launching points in time. 
}
\label{fig:Mstart}
\end{figure}

Two simulations are performed starting the outflow {\em before} disk formation, one simulation with a protostellar outflow forming shortly {\em after} disk formation (the ``fiducial run''), and a fourth simulation with a relatively late launching, but still before the onset of strong radiation pressure feedback. The most important result of this simulation series is: The resulting stellar mass is almost completely independent of when the outflow starts.
The differences among the individual simulations to the mean final value of $M_* = 31.25 \mbox{ M}_\odot$ are below 1\%.

Fig.~\ref{fig:Mstart} displays the resulting accretion histories.
In each case, the launching of the protostellar outflow immediately lowers the accretion rate.
The total momentum injected into the surroundings is larger for cases of earlier protostellar outflows.
However, once radiative feedback of the forming star becomes important, the bipolar dynamics are dominated by radiative forces.
The runs with earlier outflows have somewhat {\em higher} accretion rates during the phase $17 \le M_*/ \mbox{ M}_\odot \le 21.$
The net result is a virtually unchanged final stellar mass.
We conclude that, unless the protostellar outflow is launched into the medium after the onset of strong radiation pressure (at about $M_* \gtrapprox 20-30 \mbox{ M}_\odot$), the outflow turn-on time has little effect on the final stellar mass.

Other important details can be discerned from Fig.~\ref{fig:Mstart}.
When the infalling material hits its centrifugal barrier, i.e. when centrifugal forces balance gravity,
\vONE{
the infall halts for this parcel of gas.
Further significant growth of the stellar mass is enabled by accretion stresses which extract angular momentum from the orbiting material, either transferring it radially to the outer disk, or expelling it vertically into an outflow.  
These accretion torques in massive disks come from self-gravity-induced spiral arms.  
Such non-axisymmetric features are not treated in our 2D calculations, so we mimic the arms' effect using a shear viscosity subgrid model.  
In \citet{Kuiper:2011p21204}  we compared accretion rates from such 2D calculations against those from a 3D calculation of disk formation, finding reasonably good
agreement in the mean accretion rate for large ratios of the stress to pressure, $\alpha = 0.3 \ldots 1$.
We here set $\alpha$ to 1.  
Note that our outflow is non-rotating, and removes no orbital angular momentum.

As the infall proceeds, material with increasing specific angular momentum is stopped at ever larger radii.  
In our simulations, the accretion disk first appears once the centrifugal barrier radius exceeds the inner radius of the calculation, which occurs when $M_* \approx 7 \mbox{ M}_\odot$ here.
}
This occurrence is accompanied by a marked dip in the accretion rate for the two cases of post-disk outflow launching.
When the protostellar outflow launches before disk formation, the interaction of the outflow with high velocity infall causes oscillations of the accretion rate, which result from the positive feedback loop of how accretion depends on outflow and vice versa.
Other oscillations in the accretion rate occur when the infall rate reaches its local maximum at $M_* = 10-12 \mbox{ M}_\odot$, again due to the positive feedback loop of the outflow interacting with inflow for the cases that have both disks and outflows. 

Because the final stellar mass is virtually independent of the outflow launching time, we fix this parameter for our further studies.
We chose $M_* = 8 \mbox{ M}_\odot$, 
\vONE{i.e.~at a point in time shortly after the disk has started to form at the inner rim of the computational domain;
this choice is }
based on the implicit assumption that protostellar outflows are launched by wound-up magnetic fields at the inner rim of the accretion disk soon after its formation.

\subsection{Ratio of Ejection to Accretion Rates -- $f_{\rm ejec-acc}$ Dependence}
It is currently unclear how much of the accretion flow onto the disk is ultimately diverted into a protostellar outflow, and how the mass transfer depends on system parameters.
Most likely, the ratio $f_\mathrm{ejec-acc}$ of ejection to accretion rates will depend on several environmental properties, such as the magnetic field strength, the degree of ionization, and the accretion rate, itself.

\begin{table*}[!t]
\begin{center}
\begin{tabular}{l | c | c c c c}
Run label & 
$f_\mathrm{ejec-acc}$ [\%] & 
SFE [\%] &
$\Delta M_\mathrm{ejec} / M_\mathrm{res}$ [\%] & 
$(\Delta M_\mathrm{entr} + \Delta M_\mathrm{rad}) / M_\mathrm{res}$ [\%] &
$\Delta M_\mathrm{feedback} / M_\mathrm{res}$ [\%]
\\
\hline
A-0.01 	& ~~1	& 48.8	& ~~0.4	& 50.8 & 51.2 \\
A-0.1	& 10		& 36.1	& ~~3.1 	& 60.8 & 63.9 \\
A-0.2	& 20		& 31.2	& ~~5.8  	& 63.0 & 68.8 \\
A-0.3	& 30		& 27.9	& ~~8.5	& 63.6 & 72.1 \\
A-0.4	& 40		& 25.5	& 11.7 	& 62.8 & 74.5 \\
A-0.5	& 50		& 21.6	& 13.6	& 64.8 & 78.4
\end{tabular}
\end{center}
\caption{ 
Overview of star formation (SFE) and feedback efficiencies as function of the ratio of ejection to accretion rate $f_\mathrm{ejec-acc}$ of the protostellar outflows.
}
\label{tab:efficiencies}
\end{table*}

\begin{figure}[htbp]
\includegraphics[width=0.49\textwidth]{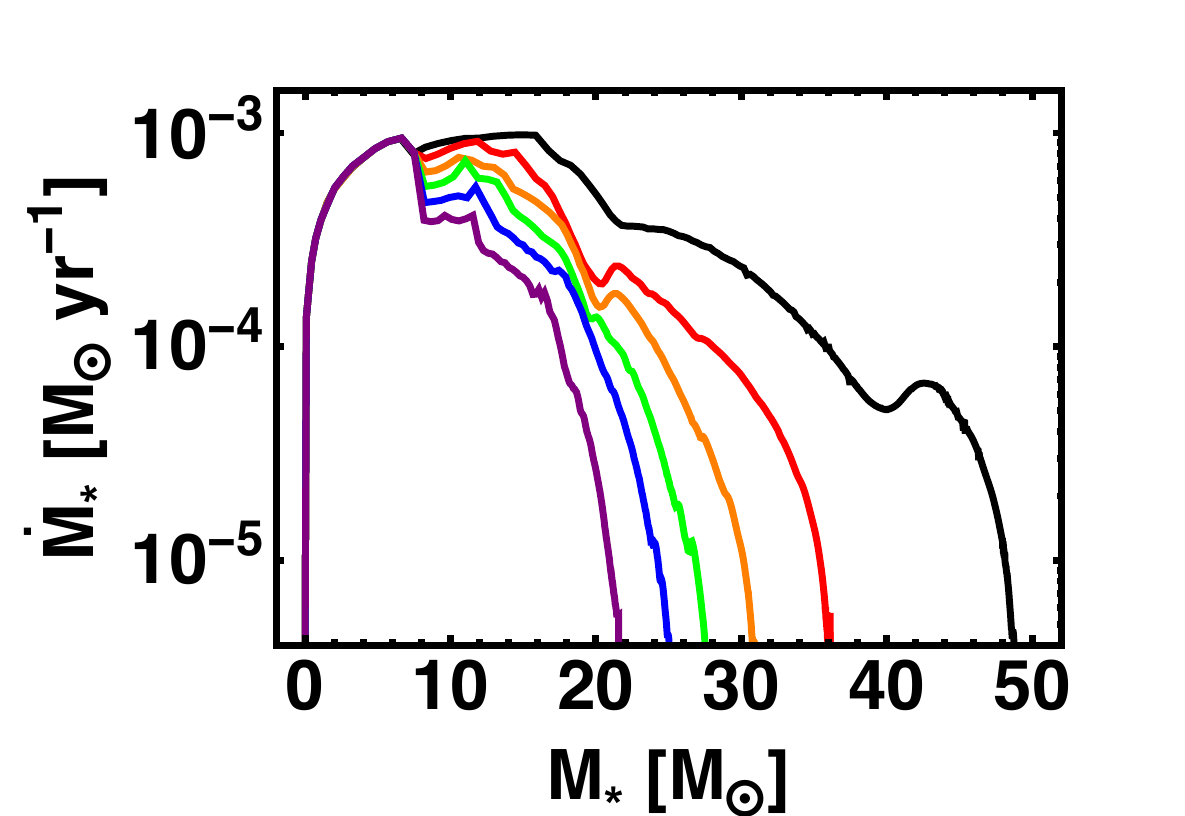}
\caption{ 
Stellar accretion rate as a function of stellar mass for six values of $f_\mathrm{ejec-acc}$, the ratio of ejection to accretion rates.
From right to left, i.e.~from black to red to orange to green to blue to purple, the curves denote the results for
$f_\mathrm{ejec-acc} = 1, 10, 20, 30, 40, \mbox{ and } 50\%$, respectively.
}
\label{fig:fmassloss}
\end{figure}

Treating $f_\mathrm{ejec-acc}$ as a free parameter, we examine its impact by varying it from 1\% up to 50\%, which should fully cover the physically most reasonable regime.
The resulting stellar accretion histories of the individual simulation runs are compared in Fig.~\ref{fig:fmassloss}.
We find that varying $f_\mathrm{ejec-acc}$ impacts the final stellar mass in three ways: reduction of the stellar accretion rate, entrainment of envelope material, and a modified impact of radiative feedback.

Whereas one can easily calculate the reduction of stellar mass growth due to the diversion of part of the inflow into outflow,
attributing the remaining differences to either entrainment or radiative feedback is less straightforward.
Qualitatively, one expects entrainment to increase with the strength of the protostellar outflow by removing mass from the stellar environment, which otherwise would be accreted by the protostar.
Reduced stellar mass growth, on the other hand, implies a lower luminosity and reduced radiative feedback.

\begin{figure}[bthp]
\begin{center}
\includegraphics[width=0.45\textwidth]{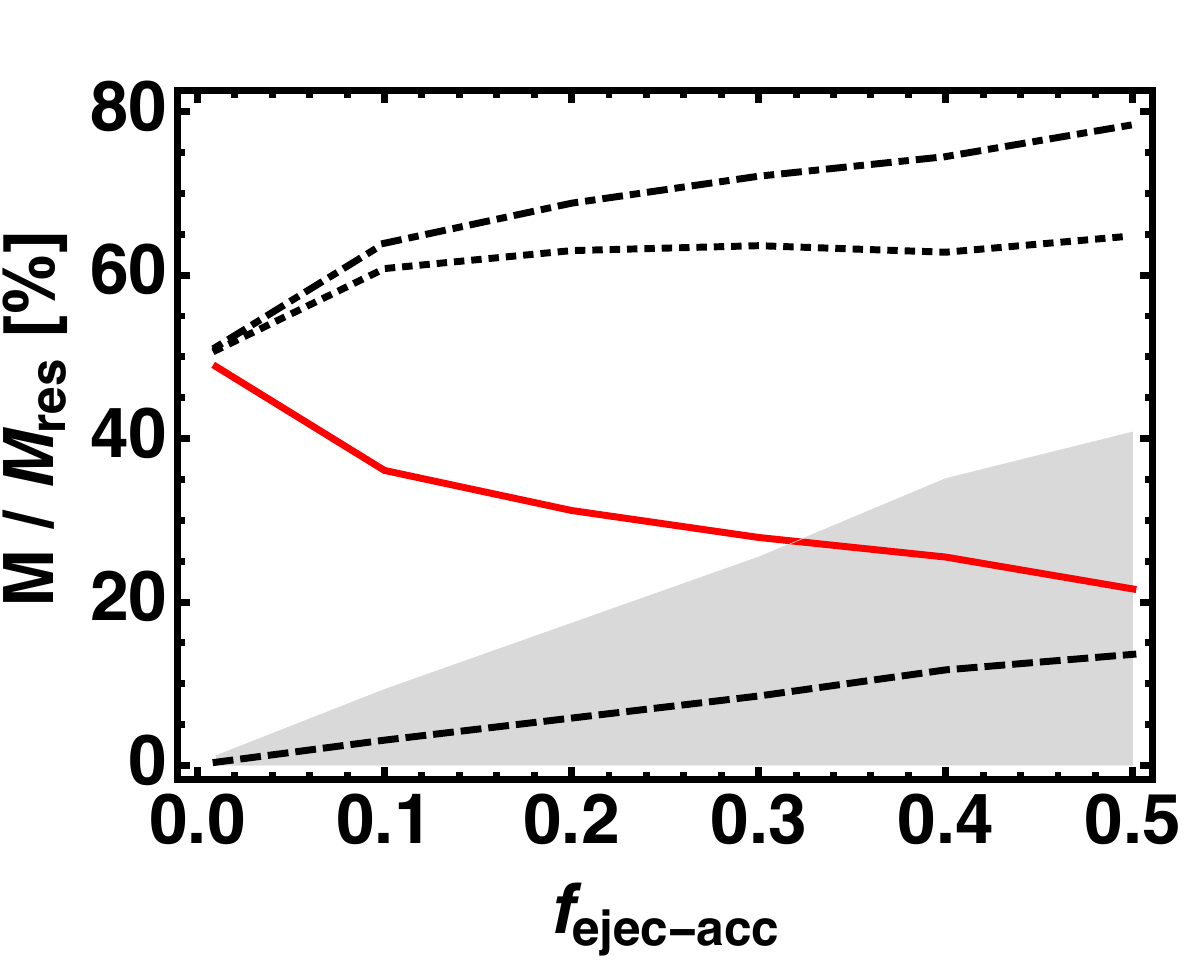}
\end{center}
\caption{
Star formation (red) and feedback efficiencies (black) as function of the ratio of ejection to accretion rates $f_\mathrm{ejec-acc}$ of the protostellar outflows.
The feedback is separated into the components: 
ejected outflow mass (lowermost dashed line), 
the combination of entrainment plus radiative mass loss of the environment (dotted line), 
the maximum level of potential entrainment (gray shaded region),
and the net impact (uppermost dot-dashed line).
}
\label{fig:feedback}
\end{figure}

An advantage of the long evolution simulations is that the net feedback can be determined quantitatively.
We stop the simulations when the mass in the computational domain is close to zero, i.e.~the initial mass of the core (here: $M_\mathrm{res} = 100 \mbox{ M}_\odot$) has either been accreted by the star
(integrated mass flux through the inner radial boundary at $10$~AU) 
or has been expelled by feedback 
(integrated flux across the outer radial boundary of the computational domain at $0.1$~pc).
More simply, the total mass loss due to feedback is simply the difference of the initial core mass and the final stellar mass: $\Delta M_\mathrm{feedback} = M_\mathrm{res} - M_*^\mathrm{final}$.

The net feedback efficiency can be further differentiated into the three individual feedback components: $\Delta M_\mathrm{feedback} = \Delta M_\mathrm{ejec} + \Delta M_\mathrm{entr} + \Delta M_\mathrm{rad}$. The first term relates to the mass ejected from the star-disk system by the jet and outflow, the second to the entrained mass in the envelope, and the last to the radiative mass loss of the reservoir.
The overall trend of the star formation efficiency and the individual feedback components is summarized in Table~\ref{tab:efficiencies} and shown in Fig.~\ref{fig:feedback}.

The mass ejected by the star-disk system can be computed from the final stellar mass $M_*^\mathrm{final}$, the launching point of the outflow $M_\mathrm{launch}$, and its ejection to accretion ratio $f_\mathrm{ejec-acc}$: 
$\Delta M_\mathrm{ejec} = \left(M_*^\mathrm{final} - M_\mathrm{launch}\right) \times f_\mathrm{ejec-acc} / (1 - f_\mathrm{ejec-acc})$.
In the simulation series of varying $f_\mathrm{ejec-acc}$, we used $M_\mathrm{launch} = 8 \mbox{ M}_\odot$.
As expected, the ejected mass increases (almost linearly) with $f_\mathrm{ejec-acc}$.

The mass loss due to entrainment and radiative feedback can be determined as the remaining difference to the total feedback effect:
$\Delta M_\mathrm{entr} + \Delta M_\mathrm{rad} = \Delta M_\mathrm{feedback} - \Delta M_\mathrm{ejec}$.
Although these two components cannot clearly be distinguished quantitatively from the simulation data, we can estimate their relative importance due to the momentum conservation of the outflowing material:
On average, the jet and outflow material is injected into the computational domain with three times the escape velocity.
This means that the entrained mass cannot be higher than three times the injected outflow mass, assuming that the entrained mass is at rest and close to the launching point.
Because the entrained mass will be gravitationally {\em infalling} prior to its entrainment, the factor three clearly denotes an upper limit.
This maximum level of entrainment is shown as a gray shaded region in Fig.~\ref{fig:feedback}.
The net mass loss is much higher than three times the injected mass, hence, the strongest feedback component for these super-Eddington stars results from their stellar radiation.

Another geometrical argument could be used to estimate an upper limit of the entrainment:
For an assumed outflow range of less than $30\degr$, measured from the symmetry axis to the cavity wall, the mass initially contained in this region is only 13~M$_\odot$.
For even the strongest outflow ($f_\mathrm{ejec-acc} = 50\%$), the radiative mass loss is still four times larger than the loss due to entrainment.

Of course, these estimates are only useful for the super-Eddington stellar mass range covered in these simulations.
For lower mass stars in the sub-Eddington regime ($M_* << 20 \mbox{ M}_\odot$), radiative forces will become less important.

As a general trend, entrainment should become more effective for higher ratios $f_\mathrm{ejec-acc}$ of ejection to accretion rates.
However, since higher values of this parameter result in less massive stars, which in turn decreases the radiative feedback,
it is difficult to distinguish quantitatively between the impact of the two effects.
Overall, entrainment plus radiation feedback increases slightly with higher $f_\mathrm{ejec-acc}$, but the net impact levels off for high values to slightly more than 60\%;
the mass loss in the most reasonable regime of outflow strengths ($f_\mathrm{ejec-acc} \ge 20\%$) is roughly constant to first order, allowing only up to 40\% of the core mass to enter the star plus disk system.
The final stellar mass then depends on how much of this mass is redirected to the jet and outflow.

Although the optically thin outflow cavity diminishes the radiative feedback (c.f.~Paper I), this influence is clearly superimposed on by the much stronger effect of redirected accretion and entrainment.
Whereas the core-scale feedback (entrainment and radiative feedback) up to 0.1~pc clearly dominates the total feedback efficiency,
the disk-scale feedback (redirection of accretion into outflow) determines the final stellar mass.
In the case of the strongest outflows ($f_\mathrm{ejec-acc} = 50\%$), the star formation efficiency drops to only 22\%.

\subsection{Wide-angle Disk Winds -- $\theta_0$ Dependence}
\begin{figure}[htbp]
\begin{center}
\includegraphics[width=0.45\textwidth]{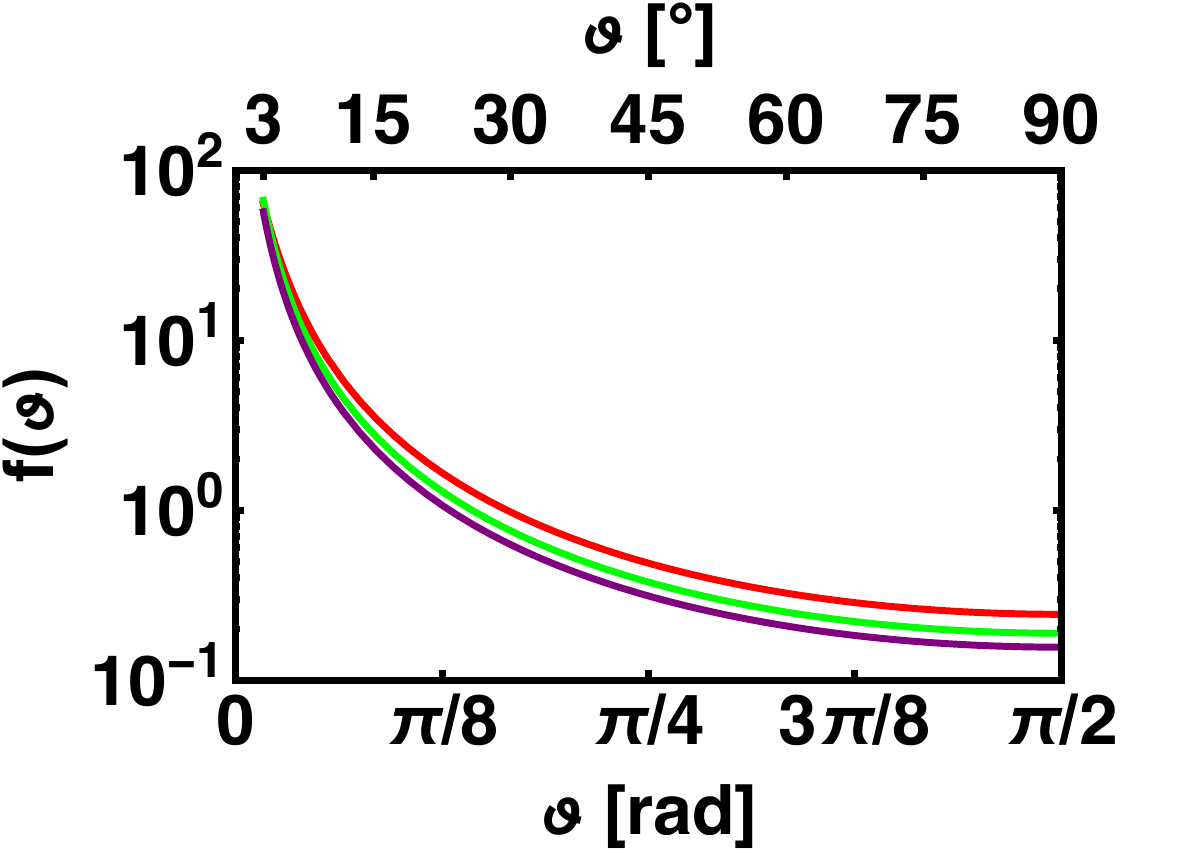}
\end{center}
\caption{
Angular distribution of the momentum flow rate of the protostellar outflow as function of the polar angle with respect to the outflow axis for the three different values of the flattening parameter: $\theta_0 = 1/30$ (red), 1/100 (green), and 1/300 (purple).
}
\label{fig:angulardependence}
\end{figure}

The jets, outflows, and winds occurring during early phases of star formation are often separated into a fast, highly collimated component and a slow wide-angle component.
The fast jet-like component prevents accretion from polar directions while creating large low density cavities, which alter the radiative force feedback during later epochs of evolution (c.f. Paper I).
The slow wide-angle component cannot prevent accretion close to the disk's high density midplane due to its insufficient momentum, 
but -- depending on its actual strength -- it could affect the disk's atmosphere and the gas at intermediate latitudes.
Here, we investigate the impact of such a disk wind by varying the strength of the slow, large-angle component of the outflow subgrid-model in three simulations, i.e. we change the flattening parameter from $\theta_0 = 1/30, 1/100$, to 1/300
(see Fig.~\ref{fig:angulardependence}).
The fast outflow component close to the poles remains the same in the three runs, but the momentum of the wide-angle disk wind component varies by a factor of 1.6 from the weakest to the strongest case.

\begin{figure}[p]
\begin{center}
\includegraphics[width=0.45\textwidth]{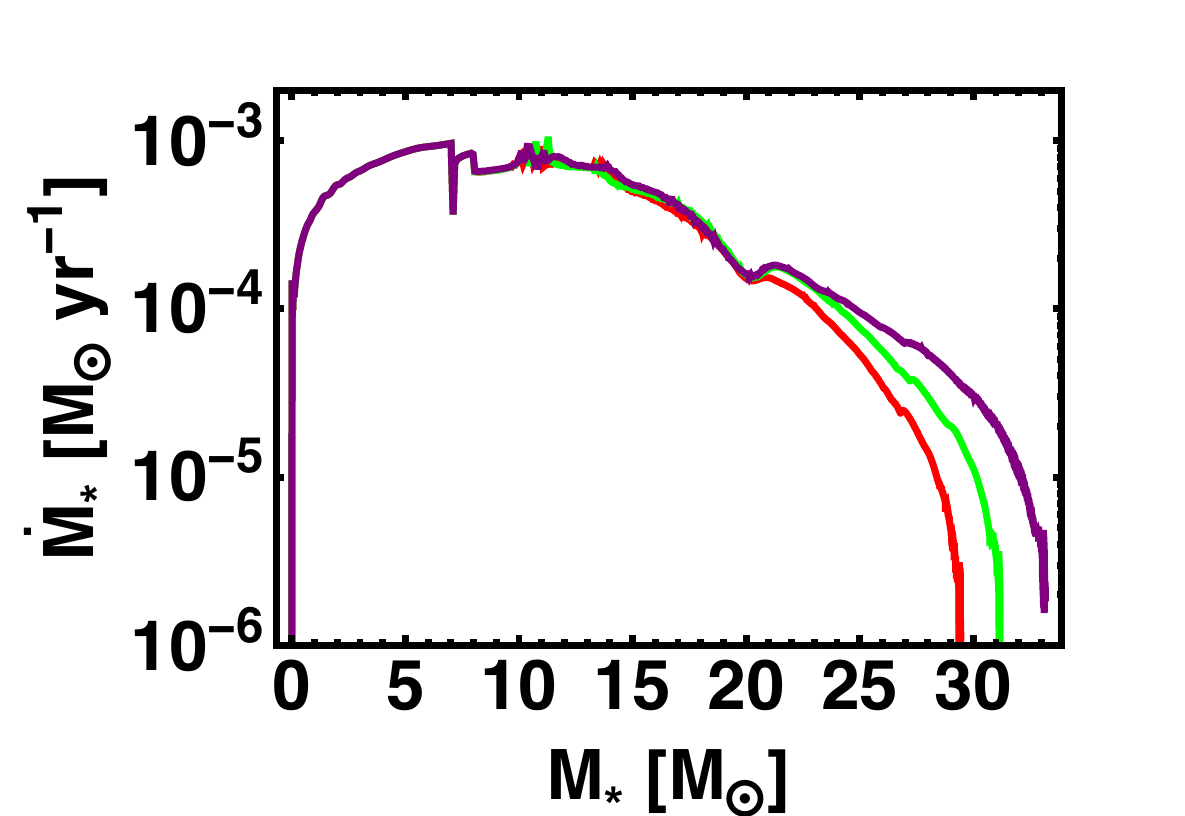}\\
\hspace{3mm}
\includegraphics[width=0.41\textwidth]{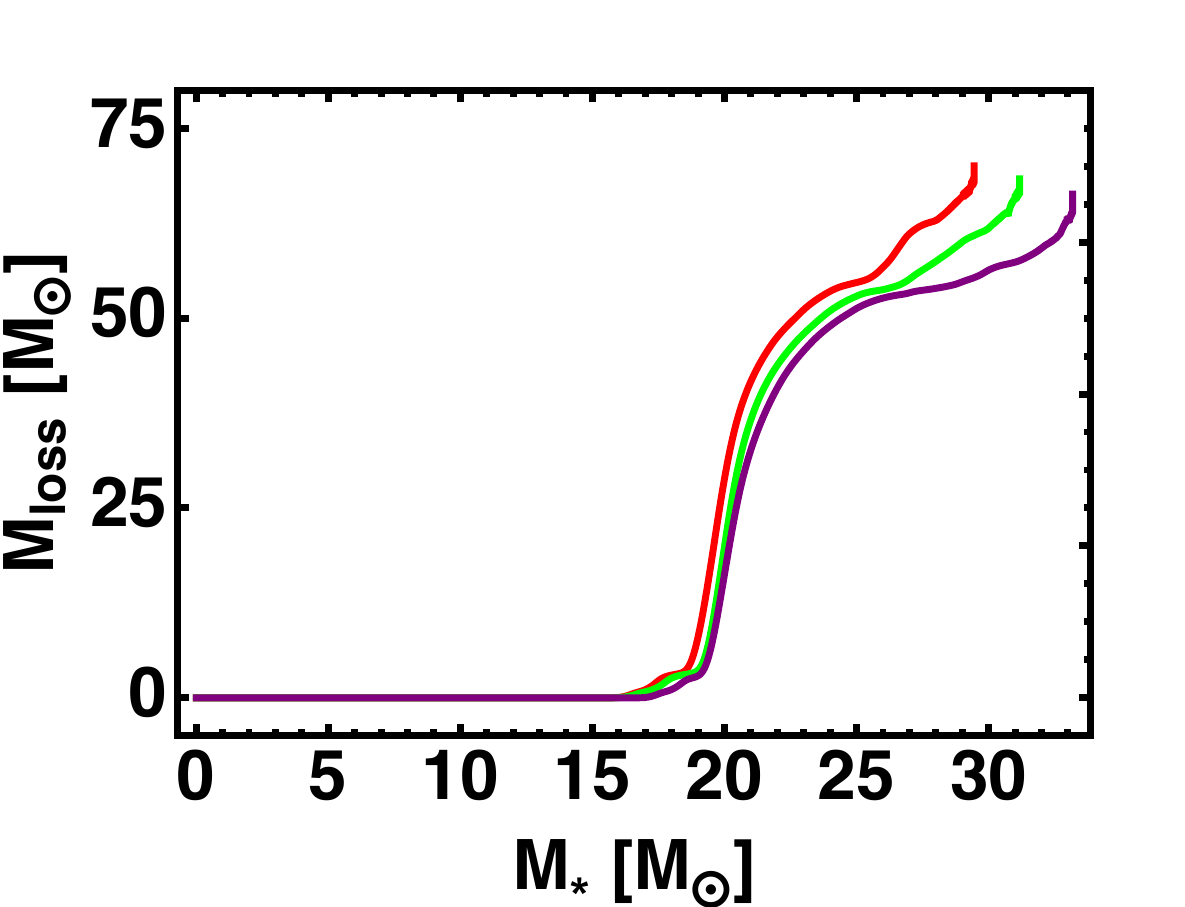}
\end{center}
\caption{
Stellar accretion rate (upper panel) and 
mass loss of the star+disk+envelope system (lower panel)
as a function of stellar mass for three different strengths of the wide-angle wind component using the same color scheme as in Fig.~\ref{fig:angulardependence}.
}
\label{fig:diskwind}
\end{figure}
\begin{figure}[p]
\hspace{-3.5mm}
\includegraphics[width=0.495\textwidth]{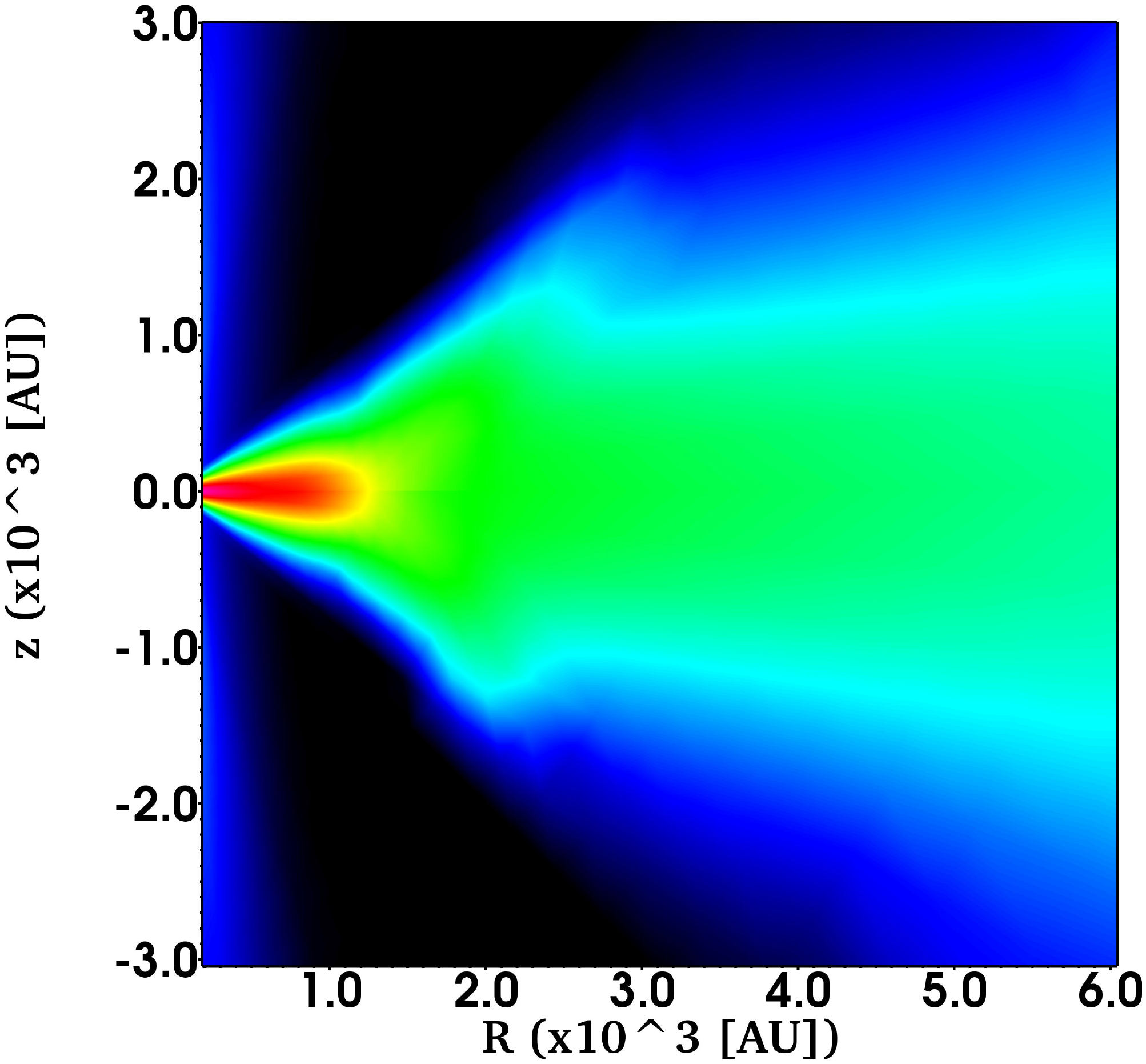}\\
\includegraphics[width=0.47\textwidth]{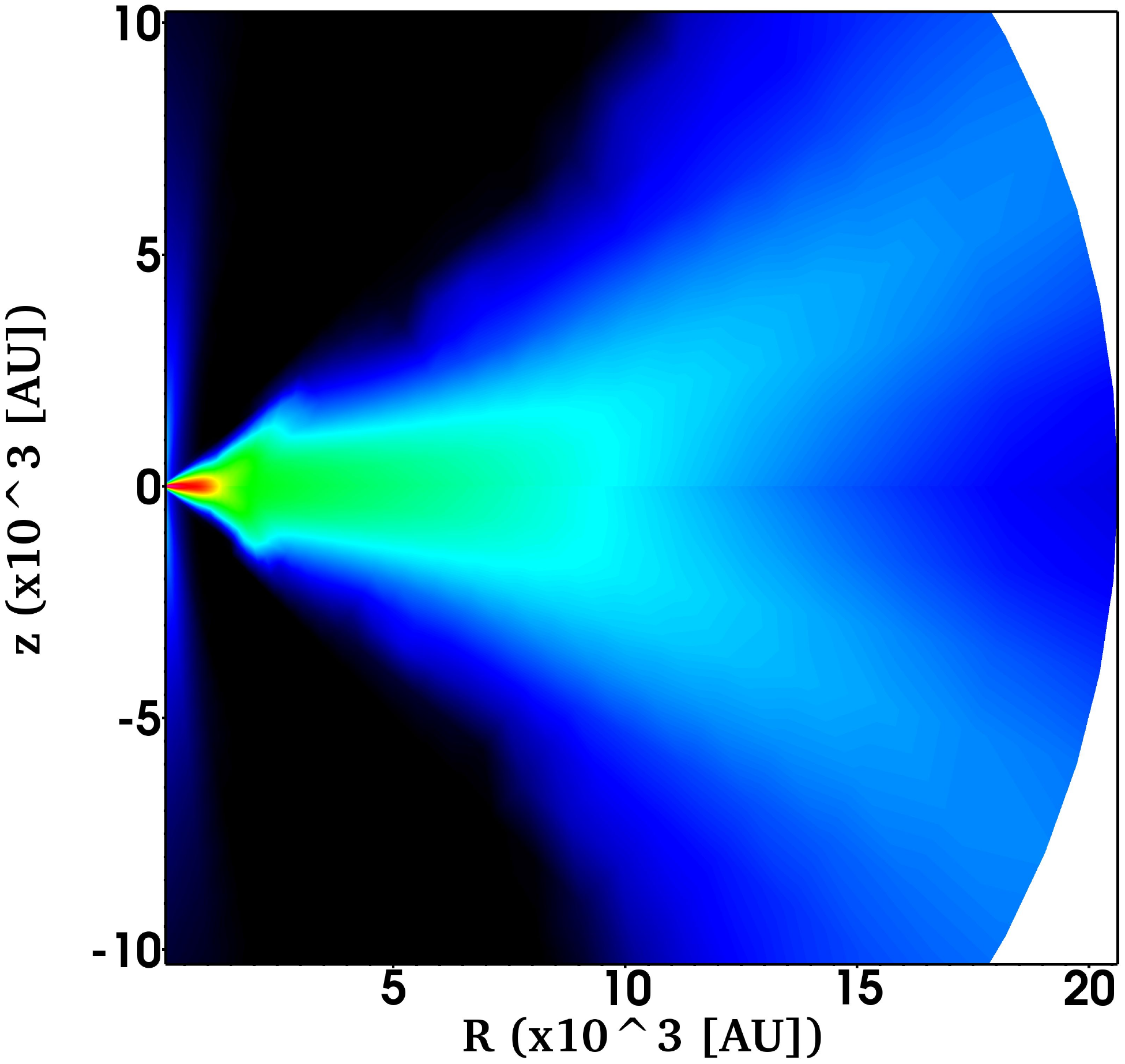}
\caption{
Comparison of gas density distributions for the cases of
strong (upper half $z>0$ of each frame) and
weak (lower half $z<0$) wide-angle disk winds at an evolutionary age
50~kyr ($M_* = 20 \mbox{ M}_\odot$).
The upper panel shows the inner $(6000 \mbox{ AU})^2$ and 
the lower panel shows the larger $(20000 \mbox{ AU})^2$ region around the forming star.
}
\label{fig:diskwind-snapshots}
\end{figure}

The resulting accretion histories of the three simulation runs are shown in Fig.~\ref{fig:diskwind}, upper panel.
The lower panel of Fig.~\ref{fig:diskwind} displays the mass loss of the star+disk+envelope system due to protostellar outflow and radiative feedback (measured as mass flow through the domain's outer boundary).
The accretion histories of the three runs are virtually identical up to about $M_* = 20 \mbox{ M}_\odot$.
After that, the star accretes less in the presence of stronger disk wind components.
Mass loss from the star+disk+envelope system for the three simulations begins to deviate after $M_* = 16 \mbox{ M}_\odot$, resulting in higher mass loss for stronger winds, as expected.
Compared to the fiducial simulation, 
the star gains $2 \mbox{ M}_\odot$ (+6\%) more with the weaker disk wind and
$1.7 \mbox{ M}_\odot$ less (-6\%) for the case of the stronger disk wind.

The wide-angle component of the disk wind reduces the amount of material accreted onto the disk due to entrainment of inter\-mediate-density gas.
Fig.~\ref{fig:diskwind-snapshots} shows the density structure of the strongest and the weakest disk wind simulations at 50~kyr ($M_* = 20 \mbox{ M}_\odot$), i.e.~after the mass loss deviates, but before noticeable differences in the stellar accretion occur.
The morphologies and dynamics of the three simulation runs are very similar at this time.
The accretion disk (flattened red structure in Fig.~\ref{fig:diskwind-snapshots}) is hardly affected by the strength of the wide-angle wind component, but
the gas available for accretion onto the disk (green region) is slightly shifted outward for the stronger wide-angle wind.
Somewhat later, the disks' evolution do differ due to the differences in large scale feeding.

We conclude that in cases of stronger disk winds, increased mass loss of the envelope due to entrainment eventually results in correspondingly lower stellar masses.
Both components of protostellar outflows, the fast, narrowly collimated jet and the wide-angle disk wind, affect the final evolution of the envelope, the accretion disk, and the protostar itself.

\subsection{Outflow broadening in time}
\label{sect:broadening}
In this section, we discuss how the opening angle of the outflow is affected by our three outflow characterization parameters.
\begin{figure}[htbp]
\begin{center}
\includegraphics[width=0.49\textwidth]{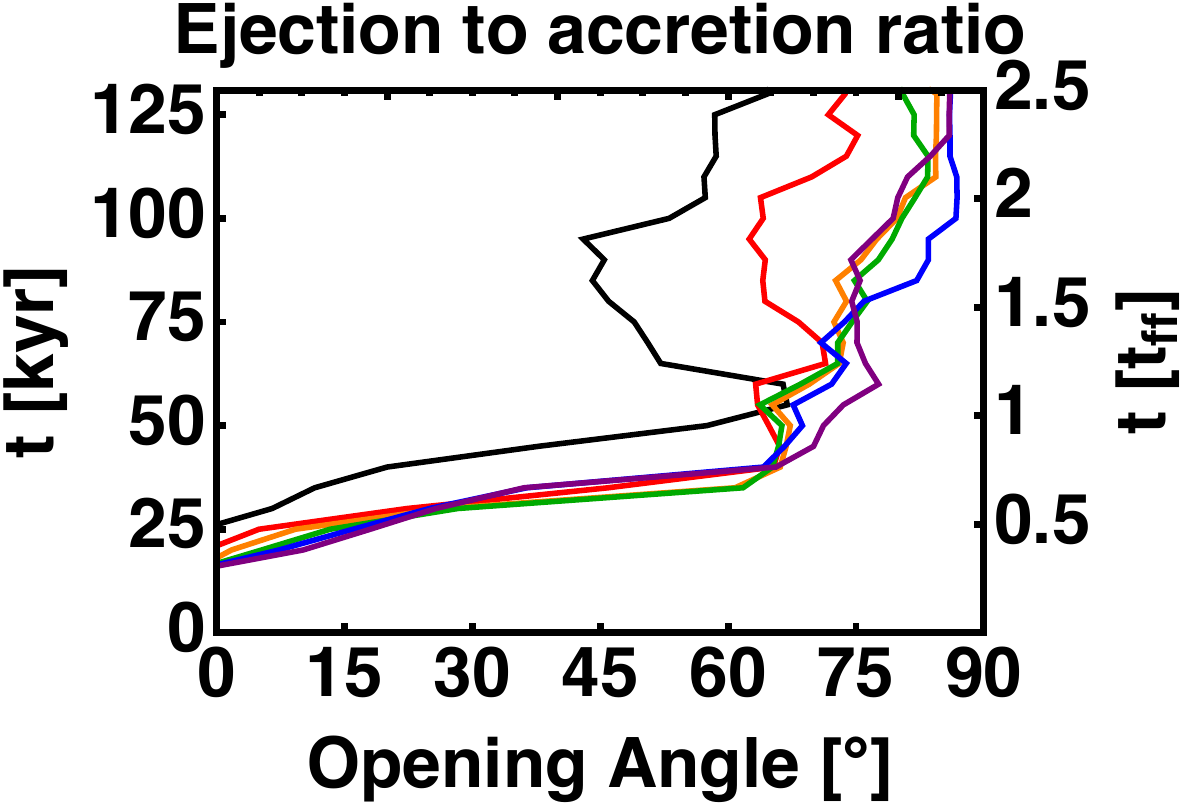}\\
\hspace{5mm}
\includegraphics[width=0.49\textwidth]{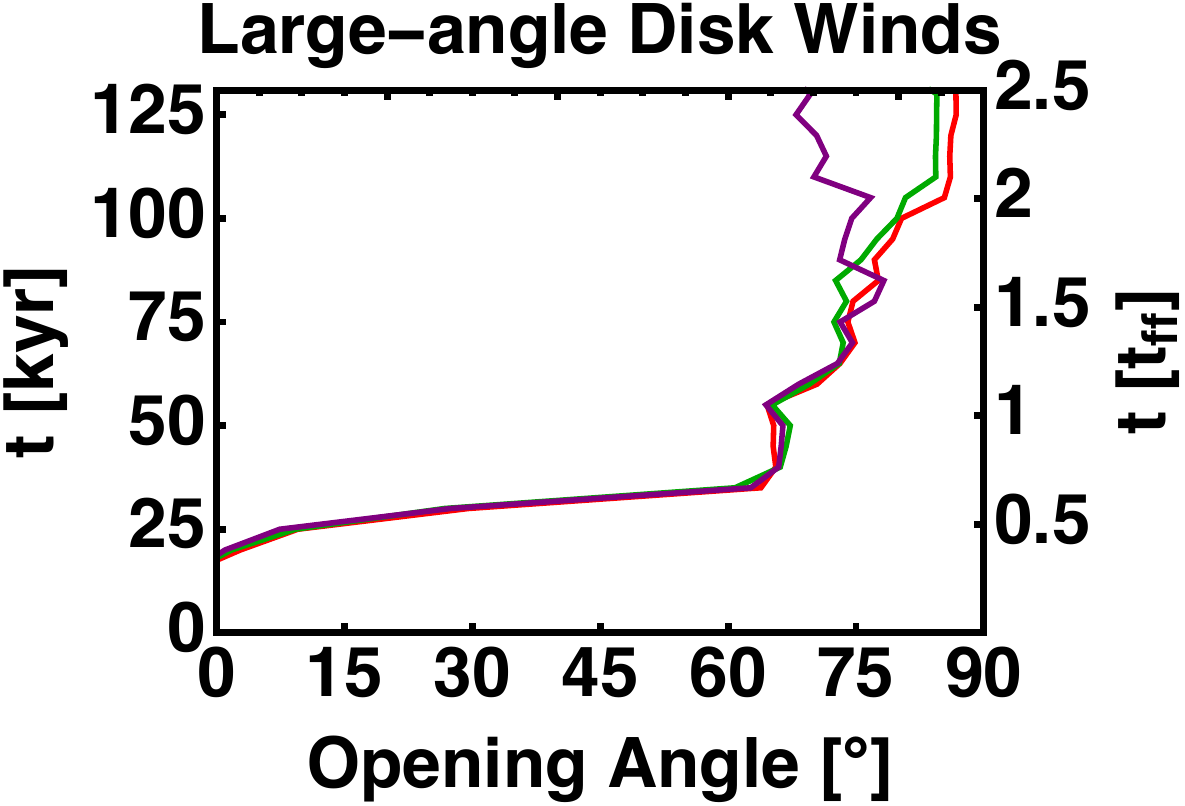}\\
\hspace{5mm}
\includegraphics[width=0.49\textwidth]{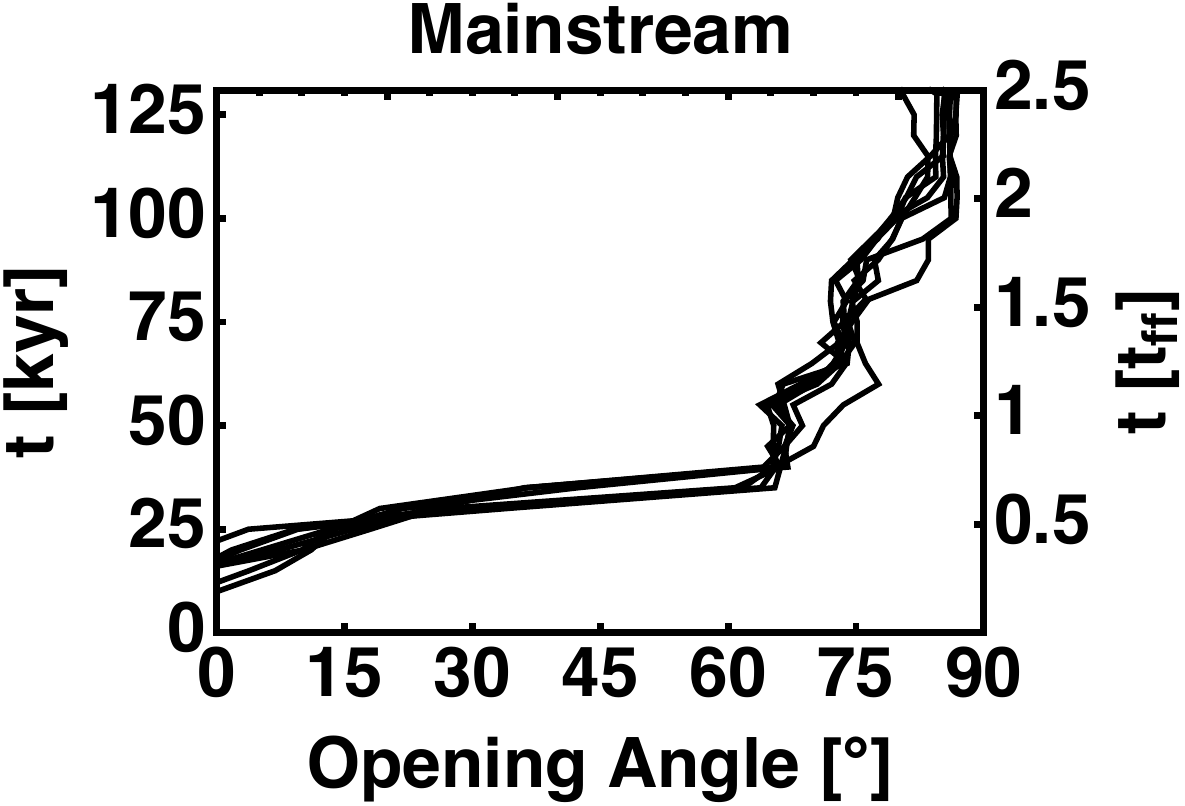}
\end{center}
\caption{
Mean opening angle $\overline{\theta_+}$ of the outflow cavities as function of time.
The upper panel shows the dependence on the ratio of ejection to accretion rates $f_\mathrm{ejec-acc}$ using the same color scheme as in \vONE{Fig.~\ref{fig:fmassloss}}.
The middle panel shows the dependence on the strength of the large angle disk wind components parameterized by $\theta_0$ using the same color scheme as in Fig.~\ref{fig:angulardependence}.
The bottom panel shows what we denote \vONE{as} the ``mainstream of outflow broadening'' (see text) and includes curves from the upper two panels selected for their similar shapes.
}
\label{fig:broadening}
\end{figure}
Because we have restricted ourselves to a specific parameterization of the outflow's launching characteristics, which are kept constant over time, we cannot determine how outflow cavities may evolve due to physical changes of the launching process.
With this caveat in mind, we shall now consider the evolution of the immediate stellar environment.
Specifically, we look at how the outflow cavity forms and then broadens due to radiation pressure feedback and the decreasing optical depth of the disk and remnant envelope.

As visualized in Fig.~\ref{fig:snapshots}, the opening angles in the simulations depend on the distance to the (proto)star.
To allow for a quantitative comparison of the different simulation runs, we focus on the time dependency of the opening angle only
and define a mean opening angle $\overline{\theta_+}$
\begin{equation}
\overline{\theta_+} = \frac{1}{R_\mathrm{max}-R_\mathrm{min}} \int_{R_\mathrm{min}}^{R_\mathrm{max}} \theta_+ ~ dr \; ,
\end{equation}
where $\theta_+(r)$ is the maximum angle of a component with positive radial velocity within the shell $[r, r+dr]$.
$R_\mathrm{min}$ and $R_\mathrm{max}$ represent the inner and outer boundaries of the computational domain at 10~AU and 0.1~pc, respectively.
Angles are measured from the symmetry axis, i.e.~a spherical wind corresponds to an opening angle of $90\degr$.

Fig.~\ref{fig:broadening} shows the outcome of this analysis for the simulation series of 
varying ejection to accretion ratios (upper panel),
varying strength of the large angle disk wind component (middle panel), and
a summary of all simulation runs (lower panel), which fall in a category that we deem ``mainstream''.
The mainstream category includes all simulations that have a similar evolution of $\overline{\theta_+}(t)$, i.e. all cases except A-0.01, A-0.1, and D-300. These exceptions have the weakest outflows.

During the early evolution ($< 25$~kyr), the outflow's opening angle does vary somewhat with the choice of $M_\mathrm{launch}$ and $f_\mathrm{ejec-acc}$, but remains below $15\degr - 20\degr$.
With the onset of strong radiation pressure, the opening angle quickly widens up to about $65\degr$.
This process takes place well within the first free-fall time of the initial core.
Subsequently, the mass in the stellar vicinity continuously declines due to accretion onto the protostar and mass loss by feedback.
As a result, the mean opening angle increases further, limiting infall and accretion to within the solid angle of the circumstellar disk (Fig.~\ref{fig:snapshots}).

At first, the opening angle of the outflow is fairly independent of the outflow configuration.
After 0.75 free-fall times ($\approx 40$~kyr), $\overline{\theta_+}$ begins to depend on the choice of outflow parameters; higher ratios of ejection to accretion rates yield broader outflows.
After 1.5 free-fall times ($\approx 80$~kyr) the strength of the large angle disk wind starts to influence the mean opening angle. 
For the weakest case (D-300) the wide-angle disk wind removes little of the disk's atmosphere and the mean opening angle begins to decrease.
Similarly, an extremely low ratio of ejection to accretion rates $f_\mathrm{ejec-acc} \le 0.1$ allows the mean opening angle to decrease after a little more than 1 free-fall time ($52.4$~kyr).
By contrast, all medium and strong outflow cases appear to be quite similar.
Stronger disk wind components or larger values of $f_\mathrm{ejec-acc}$ lead to clearing of the large angle regime between $70\degr$ and $85\degr$.

\section{Summary}
\label{sect:summary}
Following our study of protostellar outflow and radiative force feedback from massive (proto)stars in Paper I, we performed three series of parameter studies to investigate the impact of 
the time when the outflow is launched, 
the ratio of ejection to accretion rates, and
the strength of the large angle disk wind component
on the evolution of the star, its accretion disk, and large scale environment.
We focussed on the feedback efficiency, the stellar accretion rate, and the final stellar mass.

We find that in the high mass star formation regime, where the accretion is super-Eddington, the total feedback efficiency in terms of mass loss from the immediate stellar vicinity is dominated by radiative forces.
Varying e.g.~the launching time of the protostellar outflow over reasonable values does not affect the final mass of the forming star.
Even simulation runs with protostellar outflows injected well before and well after the formation of a circumstellar disk show deviations of the final stellar mass below $1\%$.

The final stellar mass is affected by the ratio of ejection to accretion rates, which
impacts the evolution of the accretion-outflow system in three ways.
First, a greater re-direction of the accretion flow into an outflow directly implies a lower stellar accretion rate.
Second, a stronger protostellar outflow is able to entrain more mass from the protostellar environment.
Third, the formation of a low density cavity alters the radiative feedback on both disk and core scales (c.f. Paper I).
A quantitative comparison of the efficiencies of the individual feedback components and the resulting final stellar masses in simulations with different ratios of ejection to accretion rates reveals
a fairly constant impact of the sum of radiative forces and entrainment independent of the ejection to accretion ratio in the regime of $f_\mathrm{ejec-acc} \ge 20 \%$.
The ejected outflow mass increases roughly linearly toward higher ratios of ejection to accretion rates.
The latter outcome implies that the efficiency of the large scale accretion flow from core to disk is only marginally influenced by the outflow.
The fact that the sum of radiative feedback and entrainment remains fairly constant for the higher values of the ratio of ejection to accretion rates implies that the decrease of radiative feedback (due to the formation of lower mass stars) is compensated by the increased entrainment of the core material.

Increasing the strength of the large angle disk wind component adds to the entrainment as it impacts the intermediate density gas at mid-latitudes between the low density outflow cavity and the high density accretion disk.
Quantitatively, with changing the relative mass flow of the large angle disk wind by $+30\%$ / $-17\%$ results in a $\pm6\%$ change of the final stellar mass.

The high mass protostars considered in this work fall in the final stellar mass range of $M_* = 20 \ldots 50 \mbox{ M}_\odot$.
For these protostars, radiative forces are the dominant feedback mechanism.
Bipolar cavities carved by protostellar outflows diminish the radiative feedback efficiency due to the ``core's flashlight effect'' (c.f.~Paper I).
Varying the outflows as we do here does not change this general conclusion. 
The new results summarized above show that the total feedback efficiency also depends on the strength of the protostellar jet and outflow, both in terms of the ratio of ejection to accretion rates and the strength of the large angle disk wind component.
While the reduction of the large scale radiative acceleration by the core's flashlight effect is quantitatively about 5\% (see Paper I), the increased mechanical feedback due to re-direction of the disk-to-star accretion flow and the entrainment of the larger-scale environment can be much stronger in the regime of reasonably strong outflows (ratios of ejection to accretion rates of more than 10\%).

Comparing the observed stellar initial mass function (IMF) and the core initial mass function suggests a core-to-star efficiency of about 30\%, especially in the case of low mass star formation.
In our study -- focussing on the high mass end of the IMF -- we find core-to-star efficiencies of 30\% for values of the ratio of ejection to accretion rates of 20-30\%.
For a much higher ratio of ejection to accretion rates of 50\%, the core-to-star efficiency is as low as 20\%.

The opening angle of the bipolar outflow cavity remains below $15\degr-20\degr$ during the early protostellar phase and then quickly opens up to $65\degr$ at the onset of radiation pressure feedback.
In the case of reasonably high ratios of ejection to accretion rates greater than 20\% and large angle disk winds that are not too weak, the outflow opening angle increases further with time.
After two free-fall times of the initial core, the outflow opening has expanded up to $85\degr$, limiting the large scale accretion flow to the shadowed regions of the circumstellar disk.

\acknowledgments
This study was conducted within the Emmy Noether research group on ``Accretion Flows and Feedback in Realistic Models of Massive Star Formation'' funded by the German Research Foundation under grant no.~KU 2849/3-1.
R.~K.~further acknowledges financial support by the German Academy of Science Leopoldina within the Leopoldina Fellowship Programme, grant No.~LPDS 2011-5, for long-term research visits at the Jet Propulsion Laboratory, CA, USA, and the University of Tokyo, Japan.
Portions of this work were conducted at the Jet Propulsion Laboratory, California Institute of Technology, operating under a contract with the National Aeronautics and Space Administration (NASA).
The authors thank Takashi Hosokawa for fruitful discussions on feedback and star formation in general.


\bibliographystyle{apj}
\bibliography{Papers1}

\end{document}